\documentstyle[preprint,aps]{revtex}
\begin{document}
\tightenlines 
\draft
\vspace{2.in}
\title{ Mixing angles and electromagnetic properties of ground
state pseudoscalar and vector meson nonets in the light-cone quark
model}  
\author{ Ho-Meoyng Choi and Chueng-Ryong Ji}
\address{
Department of Physics,
North Carolina State University,
Raleigh, N.C. 27695-8202} 
\maketitle
\narrowtext
\begin{abstract}
Both the mass spectra and the wave functions of the light pseudoscalar
($\pi,K,\eta,\eta'$) and vector($\rho,K^{*},\omega,\phi$) 
mesons are analyzed within the framework of the light-cone constituent 
quark model. A gaussian radial wave function is used as a trial
function of the variational principle for a QCD motivated Hamiltonian
which includes not only the Coulomb plus confining    
potential but also the hyperfine interaction to obtain the correct
$\rho-\pi$ splitting. For the confining potential, we use (1) harmonic
oscillator potential and (2) linear potential and compare the 
numerical results for these two cases.
The mixing angles of $\omega-\phi$ and $\eta-\eta'$ are predicted
and various physical observables such as decay constants, charge radii, 
and radiative decay rates $etc.$ are calculated. Our numerical results  
in two cases (1) and (2) are overall not much different from each other 
and have a good agreement with the available experimental data. 
\end{abstract}
\newpage
\baselineskip=20pt
\setcounter{section}{0}
\setcounter{equation}{0}
\setcounter{figure}{0}
\renewcommand{\theequation}{\mbox{1.\arabic{equation}}}
\renewcommand{\thefigure}{\mbox{1.\arabic{figure}}}
\section{Introduction }
It has been realized that the relativistic effects are crucial to describe
the low-lying hadrons made of $u,d$ and $s$ quarks and 
anti-quarks\cite{Isgur2}. The light-cone quark 
model\cite{teren,Dziem,Ji,choi,Jaus1,Jaus,Chung,choi1,Huang,schlumpf,card}
takes the advantages of the equal light-cone time($\tau=t+z/c$) 
quantization and includes the important relativistic 
effects in the hadronic wave functions. 
The distinct features of the light-cone equal-$\tau$ quantization
compared to the ordinary equal-$t$ quantization may be summarized as
the suppression of vacuum fluctuations with the decoupling of complicated
zero-modes and the conversion of the dynamical problem from boost to
rotation.

The suppression of vacuum fluctuations is due to the rational energy-momentum
dispersion relation which correlates the signs of the light-cone energy
$k^{-}=k^{0}-k^{3}$ and the light-cone momentum $k^{+}=k^{0}+k^{3}$\cite{choi}.
However, the non-trivial vacuum phenomena can still be realized in the 
light-cone quantization approach if one takes into account the 
non-trivial zero-mode($k^{+}=0$) contributions. As an example, it is
shown\cite{Rey} that the axial anomaly in the Schwinger model can be obtained
in the light-cone quantization approach by carefully analyzing the 
contributions from zero-modes. Therefore, 
in the light-cone quantization approach, one can take advantage
of the rational energy-momentum dispersion
relation and build a clean Fock state expansion of hadronic wave functions
based on a simple vacuum by decoupling the complicated non-trivial zero-modes.
The decoupling of zero-modes can be achieved in the light-cone quark model
since the constituent quark and anti-quark acquire appreciable constituent 
masses. 
Furthermore, the recent lattice QCD results\cite{Ku} indicated that the mass
difference between $\eta'$ and pseudoscalar octet mesons due to the
complicated nontrivial vacuum effect increases(or decreases) as
the quark mass $m_{q}$ decreases(or increases), $i.e.$,
the effect of the topological charge contribution should be small
as $m_{q}$ increases. This supports us to build the constituent quark
model in the light-cone quantization approach because 
the complicated nontrivial vacuum effect in QCD 
can be traded off by the rather large constituent quark masses.
One can also provide a well-established formulation of various
form factor calculations in the light-cone quantization method  
using the well-known Drell-Yan-West($q^{+}=0$) frame. 
We take this as a distinctive advantage of the light-cone quark model.
  
The conversion of the dynamical problem from boost to rotation can also
be regarded as an advantage because the rotation is compact, $i.e.$, closed
and periodic. The reason why the rotation is a dynamical 
problem in the light-cone quantization approach is because 
the quantization surface $\tau$ = 0 is not invariant 
under the transverse rotation whose direction is perpendicular to the 
direction of the quantization axis $z$ at equal $\tau$\cite{surya}.
Thus, the transverse angular momentum operator involves the interaction
that changes the particle number and it is not easy to specify
the total angular momentum of a particular hadronic state.
Also $\tau$ is not invariant under parity\cite{soper}.
We circumvent these problems of assigning the quantum numbers $J^{PC}$ to
hadrons by using the Melosh transformation of each
constituents from equal $t$ to equal $\tau$.

In our light-cone quark model of mesons, the meson state $|M>$ is thus
represented by
\begin{equation}
|M> = \Psi^{M}_{Q\bar{Q}}|Q\bar{Q}>,
\end{equation}
where $Q$ and $\bar{Q}$ are the effective dressed quark and
anti-quark. The model wave function is given by
\begin{eqnarray}
\Psi^{M}_{Q\bar{Q}}&=&
\Psi(x,{\bf k}_{\perp},\lambda_{q},\lambda_{\bar{q}})= 
\sqrt{\frac{\partial k_{n}}{\partial x}} 
\phi(x,{\bf k}_{\perp}){\cal R}(x,{\bf k}_{\perp},
\lambda_{q},\lambda_{\bar{q}}),
\end{eqnarray}
where $\phi(x,{\bf k}_{\perp})$ is the radial wave function,
$\partial k_{n}/\partial x$ is a Jacobi factor and 
${\cal R}(x,{\bf k}_{\perp},\lambda_{q},\lambda_{\bar{q}})$ is the
spin-orbit wave function obtained by the interaction-independent
Melosh transformation. When the longitudinal
component $k_{n}$ is defined by $k_{n}= (x-1/2)M_{0} +
(m^{2}_{\bar{q}}-m^{2}_{q})/2M_{0}$, the Jacobian of the variable 
transformation $\{x,{\bf k}_{\perp}\}\rightarrow {\bf k}=
(k_{n},{\bf k}_{\perp})$ is given by   
\begin{eqnarray}
\frac{\partial k_{n}}{\partial x}
&=& \frac{M_{0}}{4x(1-x)}\biggl\{ 1 - \biggl[\frac{(m^{2}_{q}
- m^{2}_{\bar{q}})}{M^{2}_{0}}\biggr]^{2}\biggr\}. 
\end{eqnarray}
The explicit spin-orbit wave function of definite spin $(S,S_{z})$
can be obtained by 
\begin{eqnarray}
{\cal R}(x,{\bf k}_{\perp},\lambda_{q}\lambda_{\bar{q}})&=&
\sum_{s_{q},s_{\bar{q}}}
<\lambda_{q}|{\cal R}^{\dagger}_{{\rm M}}(x,{\bf k}_{\perp},m_{q})|s_{q}>
\nonumber\\
&\times&
<\lambda_{\bar{q}}|{\cal R}^{\dagger}_{{\rm M}}(1-x,-{\bf k}_{\perp},
m_{\bar{q}})|s_{\bar{q}}> <\frac{1}{2}s_{q}\frac{1}{2}s_{\bar{q}}|SS_{z}>,
\end{eqnarray}
where the Melosh transformation is given by  
\begin{eqnarray}
{\cal R}_{{\rm M}}(x,{\bf k}_{\perp},m)&=&
\frac{m + xM_{0} - i\sigma\cdot(\hat{\bf n}
\times \hat{\bf k})}{\sqrt{(m+xM_{0})^{2} + {\bf k}^{2}_{\perp}}}
\end{eqnarray}
with $\hat{\bf n}=(0,0,1)$ being a unit vector in the $z$-direction.

While the spin-orbit wave function is in principle uniquely determined 
by the Melosh transformation given by Eq.(1.5), a couple of 
different schemes for handling the meson mass $M_{0}$ in Eq.(1.5) have
appeared in the literatures 
\cite{teren,Dziem,Ji,choi,Jaus1,Jaus,Chung,choi1,Huang,schlumpf,card}.  
While in the invariant meson mass
scheme\cite{teren,Jaus1,Jaus,Chung,choi1,Huang,schlumpf,card},
the meson mass square $M^{2}_{0}$ is given by
\begin{eqnarray}
M^{2}_{0}&=& \frac{{\bf k}_{\perp}^{2} + m^{2}_{q}}{x}
+ \frac{{\bf k}_{\perp}^{2} + m^{2}_{\bar{q}}}{1-x}, 
\end{eqnarray}
in the spin-averaged meson mass scheme\cite{Dziem,Ji,choi}, $M_{0}$ was taken
as the average of physical masses with appropriate weighting factors from the
spin degrees of freedom. 
Nevertheless, once the best fit parameters were used\cite{choi,choi1},  
both schemes provided the predictions that were not only pretty     
similar with each other but also remarkably good\cite{choi,Jaus}
compared to the available experimental data\cite{data} for form factors,
decay constants, charge radii $etc.$ of various light
pseudoscalar($\pi,K,\eta,\eta'$) and vector($\rho,K^{*},\omega,\phi$)
mesons as well as their radiative decay widths.
The main difference in the best fit parameters was the constituent quark
masses, $i.e.$, $m_{u}=m_{d}$= 330 MeV, $m_{s}$= 450 MeV in the
spin-averaged meson mass scheme\cite{Dziem,Ji,choi} while
$m_{u}=m_{d}$= 250 MeV, $m_{s}$= 370 MeV in the invariant meson mass
scheme\cite{Jaus}. 

Also, among the literatures\cite{Jaus,Chung,Huang} using the invariant 
meson mass scheme, some literatures\cite{Jaus,Chung} used the Jacobi 
factor $\partial k_{n}/\partial x$ in Eq.(1.2) while some\cite{Huang} 
did not. However, we have recently observed\cite{choi1} that 
the numerical results
of various physical observables from Refs.\cite{Jaus,Chung}
were almost equivalent to those of Ref.\cite{Huang}
regardless of the presence-absence of the Jacobi factor 
if the same form of radial wave function($e.g.$ Gaussian) 
was chosen and the best fit model parameters in the radial 
wave function were used.

However, the effect from the difference in the choice of 
radial wave function, $e.g.$, harmonic oscillator 
wave function\cite{Jaus,Chung,Huang} versus 
power-law wave function\cite{schlumpf}, was so substantial 
that one could not get the similar result by simply changing the
model parameters in the chosen radial wave function.  
For example, in the phenomenology of various meson radiative decays 
at low $Q^{2}$, we observed\cite{choi1} that the gaussian type wave 
function was clearly better than the power-law wave function in  
comparison with the available experimental data. On the other hand,  
the radial function so far has been mostly taken as a model wave 
function rather than as a solution of QCD motivated dynamical
equation. Even though the authors in Ref.\cite{card} adopted the quark 
potential model developed by Godfrey and Isgur\cite{Isgur2} to 
reproduce the meson mass spectra, their model predictions included 
neither the mixing angles of $\omega-\phi$ and $\eta-\eta'$ nor the  
form factors for various radiative decay processes of pseudoscalar 
and vector mesons.

In this work, we are not taking exactly the same quark potential
developed by Godfrey and Isgur\cite{Isgur2}. However, we attempt to 
fill this gap between the model wave function and the QCD motivated 
potential, which includes not only the Coulomb plus confining potential 
but also the hyperfine interaction to obtain the correct $\rho-\pi$ 
splitting. For the confining potential, we take (1) harmonic 
oscillator(HO) potential and (2) linear potential and compare the numerical 
results for these two cases. We use the variational principle to solve the
equation of motion. Accordingly, our analysis covers the mass spectra 
of light pseudoscalar($\pi,K,\eta,\eta'$) and 
vector($\rho,K^{*},\omega,\phi$) mesons and the mixing angles of 
$\omega-\phi$ and $\eta-\eta'$ as well as other observables such as 
charge radii, decay constants, radiative decay widths $etc.$.
We exploit the invariant meson mass scheme in this model.  
We also adopt the parametrization to
incorporate the quark-annihilation diagrams\cite{georgi,Isgur,scadron} 
mediated by gluon exchanges and the SU(3) symmetry 
breaking, $i.e.$, $m_{u(d)}\neq m_{s}$, in the determination of 
meson mixing angles. 

The paper is organized as follows: 
In Sec.II, we set up a simple QCD motivated effective Hamiltonian 
and use the gaussian radial wave function as a trial function of the
variational principle. We find the optimum values of the model 
parameters, quark masses($m_{u(d)},m_{s}$) and gaussian parameters( 
$\beta_{u\bar{u}}=\beta_{u\bar{d}}=\beta_{d\bar{d}},
\beta_{u\bar{s}},\beta_{s\bar{s}}$) for the two cases of confining
potentials (1) and (2). We also analyze the meson mass spectra and  
predict the mixing angles of $\omega-\phi$ and $\eta-\eta'$.
We adopt the formulation to incorporate the quark-annihilation
diagrams and the effect of SU(3) symmetry breaking in the meson
mixing angles. In Sec.III, we calculate the decay constants, charge radii, 
form factors and  radiative decay rates of various light pseudoscalar 
and vector mesons and discuss the numerical results of the two confining 
potentials (1) and (2) in comparison with the available 
experimental data. Summary and discussions follow in Sec.IV.
The details of fixing the model parameters and the mixing angle 
formulations are presented in Appendices A and B, respectively. 
\setcounter{equation}{0}
\setcounter{figure}{0}
\renewcommand{\theequation}{\mbox{2.\arabic{equation}}}
\renewcommand{\thefigure}{\mbox{1.\arabic{figure}}}
\section{ Model Description}
The QCD motivated effective Hamiltonian for the description of the meson 
mass spectra is given by\cite{Isgur2,card}   
\begin{eqnarray}
H_{q\bar{q}}|\Psi^{SS_{z}}_{nlm}> 
&=& \biggl[\sqrt{m_{q}^{2}+k^{2}} + \sqrt{m_{\bar{q}}^{2}+k^{2}} 
+ V_{q\bar{q}}\biggr]|\Psi^{SS_{z}}_{nlm}>,\nonumber\\
&=& \biggl[H_{0} + V_{q\bar{q}}\biggr]|\Psi^{SS_{z}}_{nlm}> =
M_{q\bar{q}}|\Psi^{SS_{z}}_{nlm}>,  
\end{eqnarray} 
where $M_{q\bar{q}}$ is the mass of the meson,  
$k^{2}={\bf k}^{2}_{\perp}+k^{2}_{n}$, and 
$|\Psi^{SS_{z}}_{nlm}>$ is the meson
wave function given in Eq.(1.2). In this work, 
we use the two interaction potentials $V_{q\bar{q}}$ for the 
pseudoscalar$(0^{-+})$ and vector($1^{--}$) mesons:(1) 
Coulomb plus harmonic oscillator(HO), and (2) Coulomb plus linear    
confining potentials. In addition, the hyperfine interaction,    
which is essential to distinguish vector from pseudoscalar mesons,
is included for both cases, viz.  
\begin{eqnarray}
V_{q\bar{q}}&=& V_{0}(r) + V_{\rm hyp}(r)\nonumber\\
&=& a + {\cal V}_{\rm conf.} - \frac{4\kappa}{3r}
+ \frac{2\vec{S}_{q}\cdot\vec{S}_{\bar{q}}}
{3m_{q}m_{\bar{q}}}\nabla^{2}V_{\rm Coul},
\end{eqnarray}
where ${\cal V}_{\rm conf.}= br[r^{2}]$ for the linear[HO]
potential and $<\vec{S}_{q}\cdot\vec{S}_{\bar{q}}>= 1/4[-3/4]$ for
vector[pseudoscalar] meson. Even though more realistic solution of 
Eq.(2.1) can be obtained by expanding the radial function 
$\phi_{n,l=0}(k^{2})$ onto a truncated set of HO basis 
states\cite{Isgur2,card}, $i.e.$, 
$\sum_{n=1}^{n_{\rm max}}c_{n}\phi_{n,0}(k^{2})$, our intention 
in this work is 
to explore only the $0^{-+}$ and $1^{--}$ ground state meson 
properties. Therefore, we use the $1S$ state 
harmonic wave function $\phi_{10}(k^{2})$ as 
a trial function of the variational principle
\begin{eqnarray}
\phi_{10}(x,{\bf k}_{\perp})&=& 
\biggl(\frac{1}{\pi^{3/2}\beta^{3}}\biggr)^{1/2}
\exp(-k^{2}/2\beta^{2}),  
\end{eqnarray}
where $\phi(x,{\bf k}_{\perp})$ is normalized according to
\begin{eqnarray}
\sum_{\nu\bar{\nu}}\int^{1}_{0}dx\int d^{2}{\bf k}_{\perp}
|\Psi^{SS_{z}}_{100}(x,{\bf k}_{\perp},\nu\bar{\nu})|^{2}
&=&\int^{1}_{0}dx\int d^{2}{\bf k}_{\perp}
\frac{\partial k_{n}}{\partial x}|\phi_{10}(x,{\bf k}_{\perp})|^{2}=1.
\end{eqnarray}
Because of this rather simple trial function, our results could be 
regarded as crude approximations. However, we note that this choice  
is consistent with the light-cone quark model wave 
function which has been quite successful in describing various meson 
properties\cite{Dziem,Ji,choi,Jaus1,Jaus,Chung,choi1,Huang}.
Furthermore, Eq.(2.3) takes the same form as the ground
state solution of the HO potential even though  
it is not the exact solution for the linear potential case.   
As we show in Appendix A, after fixing the parameters $a,b$ and
$\kappa$, the Coulomb plus HO potential 
$V_{0}(r)$ in Eq.(2.2), turns out to be very similar in the relevant 
range of potential($r\alt 2$ fm) to the Coulomb plus linear confining 
potentials[see Figs.1(a) and 1(b)] which are frequently used in the 
literatures\cite{Isgur2,card,lucha,Karl,gromes,isgw,isgw2}. 
The details of fixing the parameters of our model, $i.e.$, quark 
masses($m_{u(d)},m_{s}$), gaussian parameters($\beta_{u\bar{d}},
\beta_{u\bar{s}},\beta_{s\bar{s}}$) and potential parameters 
($a,b,\kappa$) in $V_{q\bar{q}}$ given by Eq.(2.2) are summarized 
in the Appendix A. 

Following the procedure listed in the Appendix A, our optimized model 
parameters are given in Table I. In fixing all of these parameters,
the variational principle[Eq.(A1)] plays the crucial role
for $u\bar{d},u\bar{s}$, and $s\bar{s}$ meson systems to share
the same potential parameters $(a,b,\kappa)$ regardless of their
quark-antiquark contents[see Figs.2(a) and 2(b)]. 

We also determine the mixing angles from the mass spectra of
$(\omega,\phi)$ and $(\eta,\eta')$. 
Identifying $(f_{1},f_{2})$=$(\phi,\omega)$ and $(\eta,\eta')$
for vector and pseudoscalar nonets, the physical meson states
$f_{1}$ and $f_{2}$ are given by
\begin{eqnarray}
|f_{1}> &=& -\sin\delta  |n\bar{n}> - \cos\delta  |s\bar{s}>,
\nonumber\\
|f_{2}> &=& \cos\delta |n\bar{n}>  - \sin\delta  |s\bar{s}>,
\end{eqnarray}
where $|n\bar{n}>\equiv 1/\sqrt{2}|u\bar{u} + d\bar{d}>$ and
$\delta=\theta_{SU(3)}-35.26^{\circ}$ is the mixing angle.
Taking into account SU(3) symmetry breaking and using the 
parametrization for the (mass)$^{2}$ matrix
suggested by Scadron\cite{scadron}, we obtain
\begin{eqnarray}
\tan^{2}\delta &=& \frac{(M^{2}_{f_{2}} - M^{2}_{n\bar{n}})
(M^{2}_{s\bar{s}} - M^{2}_{f_{1}})}{ (M^{2}_{f_{2}}-M^{2}_{s\bar{s}})
(M^{2}_{f_{1}}-M^{2}_{n\bar{n}})}, 
\end{eqnarray}
which is the model independent equation for any meson $q\bar{q}$ nonets. 
The details of obtaining meson mixing angles using quark-annihilation 
diagrams are summarized in the Appendix B. 
In order to predict the $\omega-\phi$ and $\eta-\eta'$ mixing angles,
we use the physical masses\cite{data} of 
$M_{f_{1}}=(m_{\phi},m_{\eta})$ and $M_{f_{2}}=(m_{\omega},m_{\eta'})$ 
as well as the masses of $M^{V}_{s\bar{s}}$= 996[952] MeV and 
$M^{P}_{s\bar{s}}$= 732[734] MeV obtained from the expectation 
value of $H_{s\bar{s}}$ in Eq.(2.1) for  
HO[linear] potential case[see Appendix A for more details].  
Our predictions for $\omega-\phi$ and $\eta-\eta'$ mixing angles 
for HO[linear] potential are  
$|\delta_{V}|\approx 4.2^{\circ}[7.8^{\circ}]$ and 
$\theta_{SU(3)}\approx -19.3^{\circ}[-19.6.^{\circ}]$, respectively. 
The used mass spectra of light pseudoscalar and vector
mesons are summarized in Table II. Since the signs of $\delta_{V}$
for $\omega-\phi$ mixing are not yet 
definite\cite{georgi,Isgur,scadron,Das,Sakurai,Coleman}
in the analysis of the quark-annihilation diagram[see Appendix B], 
we will keep both signs of $\delta_{V}$ when we compare various physical 
observables in the next section.  
\setcounter{equation}{0}
\setcounter{figure}{0}
\renewcommand{\theequation}{\mbox{3.\arabic{equation}}}
\renewcommand{\thefigure}{\mbox{1.\arabic{figure}}}
\section{Application}
In this section, we now use the optimum model parameters presented in
the previous section and calculate various physical observables; 
(1) decay constants of light pseudoscalar and vector mesons, 
(2) charge radii of pion and kaon, (3) form factors of neutral and 
charged kaons, and (4) radiative decay widths for the  
$V(P)\to P(V)\gamma$ and $P\to\gamma\gamma$ transitions.  
These observables are calculated for the two potentials(HO and linear)  
to gauge the sensitivity of our results.

Our calculation is carried out using the standard light-cone frame( 
$q^{+}=q^{0}+q^{3}=0$) with ${\bf q}_{\perp}^{2}=Q^{2}=-q^{2}$.
We think that this is a distinct advantage in the light-cone quark 
model because various form factor formulations are well established 
in the light-cone quantization method using this well-known 
Drell-Yan-West frame($q^{+}=0$).  
The charge form factor of the pseudoscalar meson can be expressed 
for the `+'-component of the current $J^{\mu}$ as follows  
\begin{eqnarray}
F(Q^{2})&=& e_{q}I(Q^{2},m_{q},m_{\bar{q}}) +
e_{\bar{q}}I(Q^{2},m_{\bar{q}},m_{q}), 
\end{eqnarray}  
where $e_{q}(e_{\bar{q}})$ is the charge of quark(anti-quark) and 
\begin{eqnarray}
I(Q^{2},m_{q},m_{\bar{q}})&=& \int^{1}_{0}dx\int d^{2}{\bf k}_{\perp}
\sqrt{\frac{\partial k_{n}}{\partial x}} \phi(x,{\bf k}_{\perp})
\sqrt{\frac{\partial k'_{n}}{\partial x}}\phi^{*}(x,{\bf k'}_{\perp})
\frac{{\cal A}^{2} + {\bf k}_{\perp}\cdot{\bf k'}_{\perp}} 
{ \sqrt{{\cal A}^{2} + {\bf k}^{2}_{\perp}}
\sqrt{{\cal A}^{2} + {\bf k'}^{2}_{\perp}} },  
\end{eqnarray}
with the definition of ${\cal A}$ and ${\bf k'}_{\perp}$ given by  
\begin{eqnarray}
{\cal A}&=& xm_{\bar{q}} + (1-x)m_{q},\hspace{.5cm}   
{\bf k'}_{\perp}={\bf k}_{\perp} + (1-x){\bf q}_{\perp}.  
\end{eqnarray} 
Subsequently, the charge radius of the meson can be calculated by
$r^{2} = -6dF(Q^{2})/dQ^{2}|_{Q^{2}=0}$.
Also, the decay constant $f_{P}$ of the pseudoscalar meson($P=\pi,K$) 
is given by 
\begin{eqnarray}
f_{P}&=&\frac{\sqrt{6}}{(2\pi)^{3/2}}\int^{1}_{0}dx\int d^{2}{\bf k}_{\perp}
\sqrt{\frac{\partial k_{n}}{\partial x}}\phi(x,{\bf k}_{\perp})
\frac{ {\cal A} }{ \sqrt{{\cal A}^{2} + {\bf k}^{2}_{\perp}}  }. 
\end{eqnarray}
Since all other formulae for the physical observables such as the vector
meson decay constants $f_{V}$ of $V\to e^{+}e^{-}$, decay rates for the  
$V(P)\to P(V)\gamma$ and $P\to\gamma\gamma$ transitions have already 
been given in our previous publication\cite{choi} and also in other
references($e.g.$ Ref.\cite{Jaus}), we do not list them here again.  
The readers are recommended to look at Refs.\cite{choi}
and \cite{Jaus} for the details of unlisted formulae. 
In Fig.3, we show our numerical results of the pion form factor
for HO(solid line) and linear(dotted line) cases 
and compare with the available experimental data\cite{Bebek} up to 
$Q^{2}\sim 8$ GeV$^{2}$ region. Since our model parameters of 
$m_{u}$= 0.25 GeV and $\beta_{u\bar{u}}$= 0.3194 GeV for the 
HO case are same with the ones used 
in Refs.\cite{Jaus} and \cite{schlumpf}, our numerical result 
of the pion form factor is identical with the 
Fig.2(solid line) in Ref.\cite{schlumpf}.  
In Figs.4(a) and 4(b), we show our numerical results for the form 
factors of the charged and neutral kaons and compare with the results of 
vector model dominance(VMD)\cite{bell}, where a simple two-pole model of 
the kaon form factors was assumed, $i.e.$, 
$F_{K^{+}(K^{0})}(Q^{2})= e_{u(d)}m^{2}_{\omega}/(m^{2}_{\omega}+Q^{2}) 
+ e_{\bar{s}}m^{2}_{\phi}/(m^{2}_{\phi} + Q^{2})$.
From Figs.4(a) and 4(b), we can see that the neutral kaon form factors 
using the model parameters obtained from 
HO and linear potentials are not much different from each 
other even though the charged ones are somewhat different.  

The decay constants and charge radii of various pseudoscalar and 
vector mesons for the two potential cases are given in Table III 
and compared with experimental data\cite{data,amedolia}. 
While our optimal prediction of $\delta_{V}$ was 
$|\delta_{V}|= 4.2^{\circ}[7.8^{\circ}]$ for HO[linear] potential model, 
we displayed our results for the common $\delta_{V}$ value 
with a small variation( $i.e.$,
$|\delta_{V}|= 3.3^{\circ}\pm 1^{\circ}$) in Table III to show
the sensitivity. The results for both potentials are not
much different from each other and both results are quite comparable 
with the experimental data. The decay widths of the $V(P)\to P(V)\gamma$ 
transitions are also given for the two different potential models in 
Table IV. 
Although it is not easy to see which sign of $\delta_{V}$ for 
HO potential model is more favorable  to the 
experimental data, the positive sign of $\delta_{V}$ looks a little 
better than the negative one for the processes of
$\omega(\phi)\to\eta\gamma$ and $\eta'\to\omega\gamma$ transitions.
Especially, the overall predictions of HO potential model
with the positive $\delta_{V}$ 
seem to be in a good agreement with the experimental data. 
However, more observables should be compared with the data in 
order to give more definite answer for this sign issue of 
$\omega-\phi$ mixing angle. The overall predictions of linear potential 
model are also comparable with the experimental data even though
the large variation of the mixing angle $\delta_{V}$ should be taken
into account in this case.
 
In Table V, we show the results
of $P(=\pi,\eta,\eta')\to\gamma\gamma$ decay widths obtained from our  
two potential models with the axial anomaly plus partial conservation 
of the axial current(PCAC) relations. The predictions of 
$\eta(\eta')\to\gamma\gamma$ decay widths using PCAC are in a 
good agreement with the experimental data for both HO  
and linear potential models with $\eta-\eta'$ mixing angle,
$\theta_{SU(3)}=-19^{\circ}$. The predictions of the 
decay constants for the octet and singlet mesons, $i.e.$, $\eta_{8}$ 
and $\eta_{0}$, are $f_{8}/f_{\pi}=1.254[1.324]$  
and $f_{0}/f_{\pi}=1.127[1.162]$ MeV for HO[linear]
potential model, respectively.
Our predictions of $f_{8}$ and $f_{0}$ are not much different from the
predictions of chiral perturbation theory\cite{dono} 
reported as $f_{8}/f_{\pi}= 1.25$ and 
$f_{0}/f_{\pi}=1.04\pm 0.04$, respectively.
Another important mixing-independent quantity related to 
$f_{8}$ and $f_{0}$ is the $R$-ratio defined by 
\begin{eqnarray}
R&\equiv&\biggl[\frac{\Gamma(\eta\to\gamma\gamma)}{m^{3}_{\eta}}
+\frac{\Gamma(\eta'\to\gamma\gamma)}{m^{3}_{\eta'}}\biggr]
\frac{m^{3}_{\pi}}{\Gamma(\pi\to\gamma\gamma)}
= \frac{1}{3}\biggl(\frac{f^{2}_{\pi}}{f^{2}_{8}} 
+ 8\frac{f^{2}_{\pi}}{f^{2}_{0}}\biggr). 
\end{eqnarray}  
Our predictions, $R=2.31$ and 2.17 for HO and
linear potential model cases, respectively, are quite comparable to 
the available experimental data\cite{r1,r2}, 
$R_{\rm exp}=2.5\pm0.5(stat)\pm 0.5(syst)$. Also, the $Q^{2}$-dependent
decay rates $\Gamma_{P\gamma}(Q^{2})$ are calculated from the usual
one-loop diagram\cite{choi,Jaus} and the results are shown in  
Figs.5-7. Our results for both potential models are not only very
similar with each other but also in a remarkably good 
agreement with the experimental data\cite{cello1,cello2,tpc} up to 
$Q^{2}\sim 10$ GeV$^{2}$ region. We think that the reason 
why our model is so successful for $P\to\gamma^{*}\gamma$ transition
form factors is because the $Q^{2}$-dependence($\sim 1/Q^{2}$) is
due to the off-shell quark propagator in the one-loop diagram and
there is no angular condition\cite{choi} associated with the pseudoscalar 
meson.  
\section{Summary and Discussions}
In the light-cone quark model approach,  
we have investigated the mass spectra, mixing angles,
and other physical observables of light pseudoscalar and vector
mesons using QCD motivated potentials given by Eq.(2.2).
The variational principle for the effective Hamiltonian is crucial 
to find the optimum values of our model parameters.  
As shown in Figs.1(a) and 1(b), we noticed that both central potentials 
in Eq.(2.2) are not only very similar to each other 
but also quite close to the ISGW2\cite{isgw2} model potentials.
In Figs.1(a) and 1(b), we have also included the GI potential for comparison.
Using the physical masses of ($\omega,\phi$) and 
($\eta,\eta'$), we were able to predict the $\omega-\phi$ and 
$\eta-\eta'$ mixing angles as $|\delta_{V}|\approx 
4.2^{\circ}[7.8^{\circ}]$ and
$\theta_{SU(3)}\approx -19.3^{\circ}[-19.6^{\circ}]$ for the
HO[linear] potential model, respectively.  
We also have checked that the sensitivity of the mass spectra of 
($\omega,\phi$) to $\sim 1^{\circ}[5^{\circ}]$ variation of 
$\delta_{V}$, $i.e.$, from $\delta_{V}=4.2^{\circ}[7.8^{\circ}]$ to 
$3.3^{\circ}$ for HO[linear] potential case, 
is within $1\%[5\%]$ level.

Then, we applied our models to compute the observables
such as charge radii, decay constants, and radiative decays of
$P(V)\to V(P)\gamma^{*}$ and $P\to\gamma\gamma^{*}$.
As summarized in Tables III, IV, and V, our numerical results for these 
observables in the two cases(HO and linear)
are overall not much different from each other and are in a rather good  
agreement with the available experimental
data\cite{data}. Furthermore, our results of the $R$-ratio presented in 
Eq.(3.5) are in a good agreement with the experimental data\cite{r1,r2}.    
The $Q^{2}$ dependence of $P\to\gamma\gamma^{*}$ processes were also 
compared with the experimental data up to $Q^{2}\sim 8$ GeV$^{2}$.
The $Q^{2}$-dependence for
these processes is basically given by the off-shell quark propagator
in the one-loop diagrams.  
As shown in Figs.5-7, our results are in an excellent agreement with
the experimental data\cite{cello1,cello2,tpc}.
Both the pion and kaon form factors were also predicted 
in Figs.3 and 4, respectively. 
We believe that the success of light-cone quark model hinges upon the 
suppression of complicated zero-mode contributions from the light-cone
vacuum due to the rather large constituent quark masses. The 
well-established formulation of form factors in the Drell-Yan-West 
frame also plays an important role for our model to provide comparable 
result with the experimental data. 
Because of these successful applications
of our variational effective Hamiltonian method, the extension 
to the heavy($b$ and $c$ quark sector) pseudoscalar and vector
mesons and the $0^{++}$ scalar mesons is currently under 
consideration. 

While there have been previous light-cone quark model results on
the observables that we calculated in this work, they were based
on the approach of modeling the wavefunction rather than modeling
the potential. Our approach in this work attempting to fill the gap
between the model wavefunction and the QCD motivated potential has
not yet been explored to cover as many observables as we did in
this work.
Nevertheless, it is not yet clear which sign of $\omega-\phi$ mixing 
angle should be taken, even though the overall agreement between our 
HO potential model with the positive sign, $i.e.$, 
$\delta_{V}\sim 3.3^{\circ}$ and the 
available experimental data seems to be quite good.
If we were to choose the sign of $X$ as $X>0$ in Eq.(B4), then 
the fact that the mass difference $m_{\omega}-m_{\rho}$ is 
positive is correlated with the sign of the $\omega-\phi$  
mixing angle\cite{private}. In other words, $m_{\omega}>m_{\rho}$ 
implies $\delta_{V}>0$ from Eqs.(B3)-(B5). 
Perhaps, the precision measurement of $\phi\to\eta'\gamma$
envisioned in the future at TJNAF experiment might be helpful to
give more stringent test of $\delta_{V}$.
In any case, more observables 
should be compared with the experimental data to give more definite
assessment on this sign issue.  
\acknowledgements
We are grateful to Prof. Nathan Isgur for his careful reading of this
paper and providing useful information on the sign issue of the 
$\omega-\phi$ mixing. This work was supported by the Department of 
Energy under DE-FG02-96ER40947. The North Carolina 
Supercomputing Center and the National Energy Research Scientific 
Computer Center are also acknowledged for the grant of supercomputer time. 
\newpage
\noindent
\appendix
\setcounter{section}{0}
\setcounter{equation}{0}
\setcounter{figure}{0}
\renewcommand{\theequation}{\mbox{A\arabic{equation}}}
\section{ Fixation of the model parameters using the variational
principle}
In this Appendix A, we discuss how to fix the parameters of our model, 
$i.e.$, quark masses($m_{u},m_{s}$), gaussian parameters(
$\beta_{u\bar{u}}=\beta_{u\bar{d}}, \beta_{u\bar{s}},\beta_{s\bar{s}}$)
and potential parameters $(a,b,\kappa)$ in $V_{q\bar{q}}$ given by Eq.(2.2).
In our potential model, the $\rho-\pi$ mass splitting is obtained by
the hyperfine interaction, $V_{\rm hyp}$.

Our variational method first evaluates
$<\Psi|[H_{0}+V_{0}]|\Psi>$ with a trial function $\phi_{10}(k^{2})$
that depends on the parameters $(m,\beta)$ and varies these parameters
until the expectation value of $H_{0}+V_{0}$ is a minumum.
Once these model parameters are fixed, then,
the mass eigenvalue of each meson is obtained by
$M_{q\bar{q}}=<\Psi|[H_{0}+V_{0}]|\Psi> + <\Psi|H_{\rm hyp}|\Psi>$
\footnote{As we will see later, in our fitting of the $\rho-\pi$
splitting, the rather big mass shift due to the hyperfine interaction
is attributed to the large QCD coupling constant,
$\kappa=0.3\sim0.6$.}.
In this approach, we do not discriminate the gaussian parameter set  
${\bf\beta}=(\beta_{u\bar{u}},\beta_{u\bar{s}},\beta_{s\bar{s}})$
by the spin structure of mesons.

Let us now illustrate our detailed procedures of finding the optimized
values
of the model parameters using the variational principle:
\begin{eqnarray}
\frac{\partial<\Psi|[ H_{0}+ V_{0}]|\Psi>}{\partial\beta}&=& 0.
\end{eqnarray}
From Eqs.(2.1)-(2.2) and Eq.(A1), we obtain the following equations
for HO and linear potentials: 
\begin{eqnarray}
{\rm H.O.\hspace{.1cm}potential}&:&\hspace{.5cm}
b_{h}= \frac{\beta^{3}}{3}\biggl\{
\frac{\partial<\Psi|H_{0}|\Psi>}{\partial\beta}
- \frac{8\kappa_{h}}{3\sqrt{\pi}} \biggr\},\\
{\rm Linear\hspace{.1cm}potential}&:&\hspace{.5cm}
b_{l}= \frac{\sqrt{\pi}\beta^{2}}{2}\biggl\{
\frac{\partial<\Psi|H_{0}|\Psi>}{\partial\beta}
- \frac{8\kappa_{l}}{3\sqrt{\pi}} \biggr\}, 
\end{eqnarray}
where the subscript $h[l]$ represents the HO[linear] potential
parameters. Eqs.(A2) and (A3) imply that the variational principle reduces 
a degree of freedom in the parameter space.
Thus, we have now four parameters, $i.e.$,
$\{m_{u},\beta_{u\bar{d}},a,\kappa$(or $b)\}$. However, in order to
determine these four parameters from the two experimental values of
$\rho$ and $\pi$ masses, we need to choose two input parameters.
These two parameters should be carefully chosen.
Otherwise, even though the other two parameters
are fixed by fitting the $\rho$ and $\pi$ masses, our
predictions would be poor for other observables such as the ones
in Sec.III as well as other mass spectra. From our trial and error
type of analyses, we find that $m_{u}=0.25[0.22]$ GeV is the best
input quark mass parameter for the HO[linear] potential
among the widely used $u(d)$ quark mass, $m_{u}=0.22$ GeV\cite{Isgur2},
0.25 GeV\cite{Jaus}, and 0.33 GeV\cite{Dziem,isgw,isgw2}.
For the linear potential, the string tension $b_{l}=0.18$ GeV$^{2}$ is
well known from other quark model analyses\cite{Isgur2,isgw,isgw2}
commensurate with Regge phenomenology.
Thus, we take $m_{u}=0.22$ GeV and
$b_{l}=0.18$ GeV$^{2}$ as our input parameters for the linear potential
case. However, for the HO potential, there is no
well-known quantity corresponding to the string tension and thus we
use the parameters of $m_{u(d)}= 0.25$ GeV and
$\beta_{u\bar{d}}= 0.3194$ GeV as our input
parameters which turn out to be good values to
describe various observables of both the $\pi$ and
$\rho$ mesons for the gaussian radial wave function\cite{Jaus}.

Using Eqs.(2.1),(A2) and (A3) with the input value sets of 
(1)($m_{u}$=0.25 GeV,$\beta_{u\bar{d}}$=0.3194 GeV)
for the HO potential and (2)($m_{u}=0.22$ GeV,
$b_{l}=0.18$ GeV$^{2}$) for the linear potential, we obtain the 
following parameters from the $\rho$ and $\pi$ masses, viz.,
$<\Psi|H^{V(P)}_{u\bar{d}}|\Psi>=M^{V(P)}_{u\bar{d}}=
m_{\rho(\pi)}$($P$= Pseudoscalar and $V$= Vector): 
\begin{eqnarray}
(1)& &\hspace{.2cm}{\rm H.O.\hspace{.1cm} potential}: \hspace{.3cm}
a_{h}= -0.144\hspace{.1cm}\mbox{GeV},\hspace{.2cm}
b_{h}= 0.010\hspace{.1cm}\mbox{GeV}^{3},\hspace{.2cm}
\kappa_{h} = 0.607,\\
(2)& &\hspace{.2cm}{\rm Linear \hspace{.1cm} potential}: \hspace{.3cm}
a_{l}= -0.724\hspace{.1cm}\mbox{GeV},\hspace{.2cm}
\beta_{u\bar{d}}= 0.3659\hspace{.1cm}\mbox{GeV},\hspace{.2cm}
\kappa_{l} = 0.313.
\end{eqnarray}
As shown in Fig.1(a), it is interesting to note that our two
central potentials,
Coulomb plus HO(solid line) and Coulomb plus linear
(dotted line) potentials, are not much different from each other and
furthermore quite comparable to the Coulomb plus linear quark potential
model suggested by Scora-Isgur(ISGW2)\cite{isgw2}(long-dashed line for
$\kappa=0.3$ and dot-dashed line for $\kappa=0.6$) up to the range of
$r \alt 2$ fm. Those four potentials(H.O., Linear, and ISGW2) are also
compared with the Godfrey and Isgur(GI) potential
model\cite{Isgur2}(short-dashed line) in Fig.1(a).
The corresponding string tensions, $i.e.$,
$f_{0}(r)=-dV_{0}(r)/dr$, are also shown in Fig.1(b).

Next, among various sets of $\{m_{s},\beta_{u\bar{s}}\}$
satisfying Eqs.(A2) and (A3), we find $m_{s}$= 0.48[0.45] GeV and
$\beta_{u\bar{s}}$=0.3419[0.3886] GeV for HO[linear]
potential by fitting optimally the
masses of $K^{*}$ and $K$, $i.e.$, $M^{V(P)}_{u\bar{s}}=m_{K^{*}(K)}$.
Once the set of $\{m_{s},\beta_{u\bar{s}}\}$ is fixed, then
the parameters, $\beta_{s\bar{s}}$= 0.3681[0.4128] GeV for 
HO[linear] potential, can be obtained from Eq.(A2)[(A3)].
Subsequently, $M^{V}_{s\bar{s}}$ and
$M^{P}_{s\bar{s}}$ are predicted as 996[952] MeV and 732[734] MeV
for HO[linear] potential, respectively.
As shown in Fig.2(a)[2(b)], the solid, dotted and dot-dashed lines are
fixed by the HO[linear] potential
parameter sets of $\{m_{u},\beta_{u\bar{d}}\}$,
$\{m_{s},\beta_{u\bar{s}}\}$, and $\beta_{s\bar{s}}$, respectively,
and these three lines cross the same point in the space of $b$ and
$\kappa$ if the parameters in Table I are used.

We have also examined the sensitivity of our variational
parameters and the corresponding mass spectra
using a gaussian smearing function to weaken the singularity of
$\delta^{3}(r)$ in hyperfine interaction, viz.,
\begin{eqnarray}
\delta^{3}(r)&\to& \frac{\sigma^{3}}{\pi^{3/2}}\exp(-\sigma^{2}r^{2}).
\end{eqnarray}
By adopting the well-known cut-off
value of $\sigma=1.8$\cite{Isgur2,Capstick} and repeating the same 
optimization procedure as the contact term($i.e.$, $\delta^{3}(r)$) case, 
we obtain the following parameters\footnote{For the sensitivity check of
smearing out $\delta^{3}(r)$[Eq.(A6)], we kept
$\beta_{u\bar{d}}=0.3659$ GeV for the linear potential case given
by Eq.(A5) as an input value and checked how much $b_{l}$
changed.} for each potential: 
\begin{eqnarray}
{\rm H.O.\hspace{.1cm} potential}&:& \hspace{.5cm}
a_{h}= -0.123\hspace{.1cm}\mbox{GeV},\hspace{.2cm}
b_{h}= 9.89\times 10^{-3}\hspace{.1cm}\mbox{GeV}^{3},\hspace{.2cm}
\kappa_{h} = 0.636,\\
{\rm Linear \hspace{.1cm} potential}&:& \hspace{.5cm}
a_{l}= -0.7\hspace{.1cm}\mbox{GeV},\hspace{.2cm}
b_{l}= 0.176\hspace{.1cm}\mbox{GeV}^{2},\hspace{.2cm}
\kappa_{l} = 0.332.
\end{eqnarray}
The changes of other model parameters and mass spectra are given in
Tables I and II. As one can see in Eqs.(A7)-(A8) and Tables I-II,
the effects of smearing out $\delta^{3}(r)$ are quite small and
the smearing effects are in fact negligible for our numerical
analysis in Sec.III. 
\renewcommand{\theequation}{\mbox{B\arabic{equation}}}
\section{Mixing angles of $(\eta,\eta')$ and ($\omega,\phi$)}
In this appendix, we illustrate the mixing angles of
$(\eta,\eta')$ and ($\omega,\phi$) by adopting the formulation
to incorporate the quark-annihilation diagrams and the effect of
SU(3) symmetry breaking in the meson mixing angles.

The Eq.(2.5) satisfy the (mass)$^{2}$ eigenvalue equation
\begin{eqnarray}
{\cal M}^{2}|f_{i}>&=& M^{2}_{f_{i}}|f_{i}>\hspace{.3cm}(i=1,2).
\end{eqnarray}
Taking into account SU(3) symmetry breaking, we use the following
parametrization for ${\cal M}^{2}$ suggested by Scadron\cite{scadron}
\begin{eqnarray}
{\cal M}^{2} &=& \left(\begin{array}{cc}
M^{2}_{n\bar{n}} + 2\lambda & \sqrt{2}\lambda X \\
\sqrt{2}\lambda X & M^{2}_{s\bar{s}} + \lambda X^{2}
\end{array} \right).
\end{eqnarray}
The parameter $\lambda$ characterizes the strength of the
quark-annihilation graph which couples the $I$=0 $u\bar{u}$ state
to $I$=0 $u\bar{u},d\bar{d},s\bar{s}$ states with equal strength
in the exact SU(3) limit. The parameter $X$, however, pertains to
SU(3) symmetry breaking such that the quark-annihilation graph
factors into its flavor parts, with $\lambda$ , $\lambda X$
and $\lambda X^{2}$ for the $u\bar{u}\to u\bar{u}(d\bar{d})$,
$u\bar{u}\to s\bar{s}$(or $s\bar{s}\to u\bar{u})$, and
$s\bar{s}\to s\bar{s}$ processes, respectively.
Of course, $X\to 1$ in the SU(3) exact limit.
Also, in Eq.(B2), $M^{2}_{n\bar{n}}$ and
$M^{2}_{s\bar{s}}$ describe the masses of the corresponding mesons
in the absence of mixing.

Solving Eqs.(2.5),(B1), and (B2), we obtain Eq.(2.6) and 
\begin{eqnarray}
\lambda&=& \frac{(M^{2}_{f_{1}} - M^{2}_{n\bar{n}})(M^{2}_{f_{2}}
- M^{2}_{n\bar{n}})}{2(M^{2}_{s\bar{s}}- M^{2}_{n\bar{n}})},\\
X^{2}&=& \frac{2(M^{2}_{f_{2}} -M^{2}_{s\bar{s}})(M^{2}_{s\bar{s}}
- M^{2}_{f_{1}})}{(M^{2}_{f_{2}} - M^{2}_{n\bar{n}})(M^{2}_{f_{1}}
- M^{2}_{n\bar{n}})},\\ 
\tan2\delta&=& \frac{2\sqrt{2}\lambda X}{(M^{2}_{s\bar{s}}
- M^{2}_{n\bar{n}} + \lambda X^{2} - 2\lambda)}. 
\end{eqnarray}
The Eqs.(B3) and (B4) are identical
to the two constraints, Tr(${\cal M}^{2}$)= Tr($M^{2}_{f_{i}}$) and
det(${\cal M}^{2}$)= det($M^{2}_{f_{i}}$).
The sign of $\delta$ is fixed by the signs of the $\lambda$ and $X$
from Eq.(B5). Also, since Eq.(B2) is decoupled from the subspace of
$(u\bar{u}-d\bar{d})/\sqrt{2}$, the physical masses of $m_{\pi}$ and
$m_{\rho}$ are confirmed to be the masses of $M^{P}_{n\bar{n}}$ and
$M^{V}_{n\bar{n}}$, respectively, as we used in Sec.II to fix the
parameters $(a,b,\kappa$).

Given the fixed physical masses of $M^{P}_{n\bar{n}}=m_{\pi}$
and $M^{P}_{n\bar{n}}=m_{\rho}$ together with $M_{f_{i}}(i=1,2)$,
the magnitudes of mixing angles for $\eta-\eta'$ and $\omega-\phi$
now depend only on the masses of $M^{P}_{s\bar{s}}$
and $M^{V}_{s\bar{s}}$, respectively, from Eq.(2.5).
However, from Eqs.(B3)-(B5), one can see that the sign of mixing
angle depends on the sign of parameter $X$. While $X_{P}>0$ is
well supported by the particle data
group\cite{data}($-23^{\circ}\alt\theta^{\eta-\eta'}_{SU(3)}\alt
-10^{\circ}$),
the sign of $X_{V}$ is not yet definite at the present stage of
phenomenology. Regarding on the sign of $X_{V}$, it is interesting
to note that $\delta_{V}\approx
-3.3^{\circ}(=\theta_{SU(3)}-35.26^{\circ})$($i.e.$, $X_{V}<0$)
is favored in Refs.\cite{Jaus,Das,Sakurai,Coleman}, while
the conventional Gell-Mann-Okubo mass formula for the exact
SU(3) limit($X\to 1$) predicts $\delta_{V}\approx 0^{\circ}$
in the linear mass scheme and
$\delta_{V}\approx +3.3^{\circ}$($i.e.$, $X_{V}>0$) in the
quadratic mass scheme\cite{data}. Our predictions  
for the $\omega-\phi$ and $\eta-\eta'$ mixing angles 
are given in Sec.II. 

The corresponding results of the mixing parameters $\lambda_{V(P)}$
and $X_{V(P)}$ in Eqs.(B3) and (B4) are obtained for the 
HO[linear] potential as follows
\begin{eqnarray}
\lambda_{V}&=& 0.57[0.73]m^{2}_{\pi}\hspace{.1cm}\mbox{GeV}^{2},
\hspace{.3cm} X_{V}= \pm2.10[\pm3.08],\nonumber\\
\lambda_{P}&=& 13.5[13.3]m^{2}_{\pi}\hspace{.1cm}\mbox{GeV}^{2},
\hspace{.3cm} X_{P}= 0.84[0.85].
\end{eqnarray}
Our values of $\lambda_{V}$ and $\lambda_{P}$ for both HO  
and linear potential cases are not much different from the predictions
of Ref.\cite{scadron}. The reason why $\lambda_{V}$
is much smaller than $\lambda_{P}$, $i.e.$,
$\lambda_{P}\approx 23[18]\lambda_{V}$ in our HO[linear]
case and $\lambda_{P}\approx 18\lambda_{V}$ in Ref.\cite{scadron},
may be attributed to the fact that in the quark-annihilation graph,
the $1^{--}$ annihilation graph involves one more gluon
compared to the $0^{-+}$ annihilation graph.
This also indicates the strong departure of $\eta-\eta'$ from the ideal
mixing.
\begin{table}
\caption{Optimized quark masses $(m_{q},m_{s})$
and the gaussian parameters $\beta$ for both harmonic oscillator and
linear potentials obtained from the variational principle. 
$q$=$u$ and $d$.}
\begin{tabular}{cccccc}
Potential& $m_{q}$[GeV] & $m_{s}$[GeV] & $\beta_{q\bar{q}}$[GeV] & 
$\beta_{s\bar{s}}$[GeV]& $\beta_{q\bar{s}}$[GeV]\\
\tableline
H.O.& 0.25 & 0.48 & 0.3194 & 0.3681[0.3703]$^{[a]}$ & 0.3419[0.3428]\\
Linear&0.22 & 0.45 & 0.3659 & 0.4128[0.4132]&0.3886[0.3887] \\  
\end{tabular}
\end{table}
$^{[a]}$ The values in parentheses are results from the smearing 
function\cite{Isgur2,Capstick} in Eq.(A6) instead of the contact term. 
\begin{table}
\caption{Fit of the ground state meson masses with
the parameters given in Table I. Underline masses are input data.
The masses of $(\eta,\eta')$ and $(\omega,\phi)$ were used to determine
the mixing angles of $\eta-\eta'$ and $\omega-\phi$, respectively.}
\begin{center}
\begin{tabular}{cccccccccc}
$^{1}S_{0}$ & Experiment[MeV] & H.O.& Linear&   
$^{3}S_{1}$ & Experiment & H.O.& Linear\\ 
\hline
$\pi$ & 135$\pm$0.00035 & \underline {135}&\underline{135} 
&$\rho$ & 770$\pm$ 0.8 & \underline {770}& \underline{770}\\
$K$ & 494$\pm$ 0.016 & 470[469]$^{[b]}$& 478[478] & $K^{*}$ & 892$\pm$ 0.24 
& 875[875]& 850[850]\\  
$\eta$ & 547$\pm$ 0.19 & \underline{547}& \underline{547} 
& $\omega$ & 782$\pm$ 0.12 & \underline{782}& \underline{782} \\
$\eta'$ & 958$\pm$0.14 & \underline{958}& \underline{958} 
& $\phi$ & 1020$\pm$0.008 & \underline{1020} &\underline{1020} \\
\end{tabular}
\end{center}
\end{table}
$^{[b]}$ The values in parentheses are results from the smearing function
in Eq.(A6) instead of the contact term. 

\newpage 
\begin{table}
\caption{Decay constants and charge radii for various pseudoscalar and
vector mesons. For comparison, we use 
$|\delta_{V}|=3.3^{\circ}\pm 1^{\circ}$ for both potential cases. 
The experimental data are taken from Ref.[17], unless otherwise noted.} 
\begin{tabular}{cccccr}
&\multicolumn{2}{c}{$\delta_{V}=-3.3^{\circ}\pm 1^{\circ}$}
&\multicolumn{2}{c}{$\delta_{V}=+3.3^{\circ}\pm 1^{\circ}$}& \\ 
Observables & H.O.& Linear &H.O. & Linear  & Experiment\\  
\tableline
$f_{\pi}$ [MeV]& 92.4 & 91.8& 92.4 & 91.8 &92.4$\pm$0.25 \\ 
$f_{K}$ [MeV]& 109.3 & 114.1& 109.3 & 114.1& 113.4$\pm$ 1.1 \\   
$f_{\rho}$ [MeV]& 151.9 & 173.9 &151.9 &173.9 & 152.8$\pm$ 3.6 \\ 
$f_{K^{*}}$ [MeV]& 157.6 & 180.8& 157.6 & 180.8 & ---\\ 
$f_{\omega}$ [MeV]& 45.9$\pm$1.4 & 52.6$\pm$1.6 &
55.1$\pm$1.3 & 63.1$\pm$1.5 & 45.9$\pm$ 0.7 \\  
$f_{\phi}$ [MeV]& 82.6$\mp$ 0.8 & 94.3$\mp$0.9& 
76.7$\mp$ 1.0 & 87.6$\mp$1.1 & 79.1$\pm$ 1.3  \\ 
$r^{2}_{\pi}$ [fm$^{2}$]& 0.449 & 0.425& 0.449& 0.425
& 0.432 $\pm$ 0.016 [32]\\
$r^{2}_{K^{+}}$ [fm$^{2}$]& 0.384 & 0.354& 0.384 & 0.354
& 0.34$\pm$ 0.05 [32]\\
$r^{2}_{K^{0}}$ [fm$^{2}$]& $-0.091$ & $-0.082$ & $-0.091$ &  
$-0.082$ & $-0.054\pm$ 0.101 [32]\\ 
\end{tabular}
\end{table}
\begin{table}
\caption{Radiative decay widths for the $V(P)\to P(V)\gamma$ transitions.
The mixing angles, $\theta_{SU(3)}=-19^{\circ}$ for $\eta-\eta'$ and 
$|\delta_{V}|=3.3^{\circ}\pm 1^{\circ}$ for $\omega-\phi$, are used  
for both potential models, respectively.
The experimental data are taken from Ref.[17].} 
\label{values}
\begin{tabular}{lccccr}
&\multicolumn{2}{c}{$\delta_{V}=-3.3^{\circ}\pm 1^{\circ}$}
&\multicolumn{2}{c}{$\delta_{V}=+3.3^{\circ}\pm 1^{\circ}$}& \\ 
Widths&H.O. & Linear & H.O. & Linear & Experiment[keV]\\
\tableline
$\Gamma(\rho^{\pm}\to\pi^{\pm}\gamma)$ & 76 & 69 & 76 &69 
&$68\pm 8$ \\
$\Gamma(\omega\to\pi\gamma)$ & 730$\pm$1.3 & 667$\pm$1.3
& 730$\mp$1.3 & 667$\mp$1.3 & $717\pm 51$ \\
$\Gamma(\phi\to\pi\gamma)$ & 5.6$^{-2.9}_{+3.9}$& 5.1$^{-2.6}_{+3.6}$   
& 5.6$^{+3.9}_{-2.9}$ & 5.1$^{+3.6}_{-2.6}$ &  $5.8\pm 0.6$ \\
$\Gamma(\rho\to\eta\gamma)$ & 59& 54 & 59 & 54 &  $58\pm 10$ \\
$\Gamma(\omega\to\eta\gamma)$ & 8.7$\mp$ 0.3 & 7.9$\mp$0.3
& 6.9$\mp$ 0.3 & 6.3$\mp$0.3 & $7.0\pm 1.8$ \\
$\Gamma(\phi\to\eta\gamma)$ & 38.7$\pm$ 1.6  & 37.8$\pm$ 1.5
& 49.2$\pm$ 1.6 & 47.6$\pm$ 1.5 & $55.8\pm 3.3$\\
$\Gamma(\eta'\to\rho\gamma)$ & 68 & 62 & 68 & 62 & 61 $\pm8 $ \\
$\Gamma(\eta'\to\omega\gamma)$ & 4.9$\pm$ 0.4 & 4.5$\pm$ 0.4
& 7.6$\pm$ 0.4 & 7.0$\pm$ 0.4 & $6.1\pm 1.1$ \\
$\Gamma(\phi\to\eta'\gamma)$ & 0.41$\mp$0.01& 0.39$\mp$0.01 
&0.36$\mp$ 0.01 & 0.34$\mp$0.01 & $< 1.8$\\
$\Gamma(K^{*0}\to K^{0}\gamma)$ & 124.5 & 116.6& 124.5& 116.6& 117$\pm$ 10\\
$\Gamma(K^{*+}\to K^{+}\gamma)$ & 79.5 & 71.4& 79.5 & 71.4 & 50 $\pm$ 5 \\ 
\end{tabular}
\end{table}
\begin{table}
\caption{Radiative decay widths $\Gamma(P\to\gamma\gamma)$ 
obtained by using the axial anomaly plus PCAC relations. 
$\theta_{SU(3)}=-19^{\circ}$ for $\eta-\eta'$ mixing 
is used for both potential cases.  
The experimental data are taken from Ref.[17].}
\begin{tabular}{ccccr}
Widths &  H.O. & Linear & Experiment\\
\tableline
$\Gamma(\pi\to\gamma\gamma)$ &  7.73& 7.84 & $7.8\pm 0.5$[eV]\\
$\Gamma(\eta\to\gamma\gamma)$ & 0.42& 0.42 & $0.47\pm0.0 5$[keV]\\
$\Gamma(\eta'\to\gamma\gamma)$ & 4.1& 3.9 & $4.3\pm 0.6$[keV] \\
\end{tabular}
\end{table}

\figure{\hspace{.2in} Fig.1(a). 
The central potential $V_{0}(r)$ versus $r$.
Our Coulomb plus harmonic oscillator(solid line) and linear(dotted) 
potentials are compared with the quasi-relativistic  
potential of ISGW2\cite{isgw2} model with $\kappa=0.3$(long-dashed line) 
and $\kappa=0.6$(dot-dashed line) and the relativized
potential of GI\cite{Isgur2} model(short-dashed line).}
\figure{\hspace{.2in} Fig.1(b). The central force $f_{0}(r)$ versus
$r$. Our force for the linear potential is the same 
as that of ISGW2 with $\kappa=0.3$(dotted lines). 
The forces of GI and ISGW2 with $\kappa=0.6$ are the same with
each other(dashed lines). Our force for the harmonic oscillator
potential(solid line) are quite comparable with the other four forces 
up to the range of $r\alt 2$ fm.}     
\figure{\hspace{.2in} Fig.2(a). The variational principle 
satisfying Eq.(A2). The solid, dotted, and dot-dashed lines 
are fixed by the sets of ($m_{u},\beta_{u\bar{u}}$),   
($m_{s},\beta_{u\bar{s}}$), and ($m_{s},\beta_{s\bar{s}}$), 
respectively.} 
\figure{\hspace{.2in} Fig.2(b). The variational principle
satisfying Eq.(A3). The same line codes are used as Fig.2(a).}  
\figure{\hspace{.2in} Fig.3. The charge form factor for the pion 
compared with data taken from Ref.\cite{Bebek}. The solid and dotted
lines correspond to the results of harmonic oscillator and 
linear potential cases, respectively.} 
\figure{\hspace{.2in} Fig.4(a). Theoretical predictions of
charged $K^{+}$ form factors using the parameters of both harmonic
oscillator(solid) and linear(dotted) potentials compared with a simple  
two-pole VMD model\cite{bell}(dot-dashed), $F^{\rm VDM}_{K^{+}(K^{0})}
= e_{u(d)}m^{2}_{\omega}/(m^{2}_{\omega} + Q^{2}) 
+ e_{\bar{s}}m^{2}_{\phi}/(m^{2}_{\phi} + Q^{2})$.}
\figure{\hspace{.2in} Fig.4(b). Theoretical predictions of
neutral $K^{0}$ form factors. 
The same line codes are used as Fig.4(a).} 
\figure{\hspace{.2in} Fig.5. The decay rate for the 
$\pi\to\gamma^{*}\gamma$  transition obtained from 
the one-loop diagram. 
Data are taken from Refs.\cite{cello1,cello2}.} 
\figure{\hspace{.2in} Fig.6. The decay rate for the 
$\eta\to\gamma^{*}\gamma$ transition obtained from 
the one-loop diagram. 
Data are taken from Refs.\cite{cello1,cello2,tpc}.} 
\figure{\hspace{.2in} Fig.7. The decay rate for the 
$\eta'\to\gamma^{*}\gamma$ transition obtained from 
the one-loop diagram. 
Data are taken from Refs.\cite{cello1,cello2,tpc}.} 

\begin{references}
\bibitem{Isgur2} S. Godfrey and N. Isgur, Phys. Rev. D {\bf 32}, 189(1985).  
\bibitem{teren} M. V. Terent'ev, Yad. Fiz. {\bf 24}, 207(1976) [ Sov.J. Nucl.
Phys. {\bf 24}, 106(1976)]; V. B. Berestetsky and M. V. Terent'ev, $ibid$.
{\bf 24}, 1044(1976) [ {\bf 24}, 547(1976)]; {\bf 25}, 653(1977)[ {\bf 25},
347(1977)].  
\bibitem{Dziem} Z. Dziembowsky and L. Mankiewicz, Phys. Rev. Lett. {\bf 58},
2175(1987); Z. Dziembowsky, Phys. Rev. D {\bf 37}, 778(1988).
\bibitem{Ji} C.-R. Ji and S.R. Cotanch, Phys. Rev. D {\bf 41}, 2319(1990);
C.-R. Ji, P.L. Chung and S.R. Cotanch, Phys. Rev. D {\bf 45}, 4214(1992). 
\bibitem{choi} H.-M. Choi and C.-R.Ji, Nucl. Phys. A {\bf 618}, 291(1997). 
\bibitem{Jaus1} W. Jaus, Phys. Rev. D {\bf 41}, 3394(1990). 
\bibitem{Jaus} W. Jaus, Phys. Rev. D {\bf 44}, 2851(1991). 
\bibitem{Chung} P.L. Chung, F. Coester, and W.N. Polyzou, Phys. Lett. 
B {\bf 205}, 545(1988). 
\bibitem{choi1} H.-M. Choi and C.-R.Ji, Phys. Rev. D {\bf 56}, 6010(1997).  
\bibitem{Huang} T.Huang,B.-Q. Ma,and Q.-X.Shen, Phys.Rev.D {\bf 49},
1490(1994).
\bibitem{schlumpf} F.Schlumpf, Phys.Rev.D.{\bf 50}, 6895(1994). 
\bibitem{card} F. Cardarelli {\em et al.}, Phys. Lett. B {\bf 349}, 
393(1995); {\bf 359}, 1(1995); {\bf 332}, 1(1994). 
\bibitem{Rey} C.-R. Ji and S.J. Rey, Phys. Rev. D {\bf 53}, 5815(1996). 
\bibitem{Ku} Y. Kuramashi {\em et al.}, Phys. Rev. Lett. {\bf 72}, 3448(1994). 
\bibitem{surya} C.-R.Ji and Y.Surya, Phys.Rev. D {\bf 46}, 3565(1992). 
\bibitem{soper} D.E.Soper, ph.D. Thesis, SLAC Report No. 137 (1971). 
\bibitem{data} Particle Data Group, R. M. Barnett {\em et al.},
Phys. Rev. D {\bf 54}, 1(1996). 
\bibitem{georgi} A. De R\'{u}jula, H. Georgi, and S. Glashow,
Phys. Rev. D {\bf 12}, 147(1975).
\bibitem{Isgur} N. Isgur, Phys. Rev. D {\bf 12}, 3770(1975);
{\bf 13}, 122(1976). 
\bibitem{scadron} M. D. Scadron, Phys. Rev. D {\bf 29}, 2076(1984). 
\bibitem{lucha} W.Lucha, F.F.Sch\"{o}berl, and D. Gromes,
Phys. Rep. {\bf 200}, 127(1991).
\bibitem{Karl} N. Isgur and G. Karl, Phys. Lett. B {\bf 72},
109(1977).
\bibitem{gromes} D. Gromes and I. O. Stamatescu, Nucl. Phys. B {\bf 112},
213(1976).
\bibitem{isgw} N. Isgur, D. Scora, B. Grinstein, and M. B. Wise,
Phys. Rev. D {\bf 39}, 799(1989).
\bibitem{isgw2} D. Scora and N. Isgur, Phys. Rev. D {\bf 52}, 2783(1992).
\bibitem{Capstick} S. Capstick and N. Isgur, Phys. Rev. D{\bf 34},
2809(1986).
\bibitem{Das} T. Das, V. S. Mathur, and S. Okubo, Phys. Rev. Lett. {\bf 19},
470(1967); J. J. Sakurai, $ibid$. {\bf 19}, 803(1967).
\bibitem{Sakurai} R. J. Oakes and J. J. Sakurai, Phys. Rev. Lett. {\bf 19},
1266(1967).
\bibitem{Coleman} S. Coleman and H. J. Schnitzer, Phys. Rev. {\bf 134},
B863(1964); N. M. Kroll, T. D. Lee, and B. Zumino, $ibid$. {\bf 157},
1376(1967).
\bibitem{Bebek} C. J. Bebek $et al$., Phys. Rev. D {\bf 17}, 1693(1978). 
\bibitem{bell} J. J. Sakurai, K. Schilcher and  M. D. Tran, Phys. Lett.
B {\bf 102}, 55(1981); J. S. Bell and J. Pasupathy, {\em ibid.} B {\bf 83}, 
389(1970).    
\bibitem{amedolia} R. A. Amendolia {\em et al.}, Phys. Lett. B {\bf 178},
435(1986).
\bibitem{dono} J. F. Donoghue, B. R. Holstein, and Y. C. R. Lin,
Phys. Rev. Lett. {\bf 55}, 2766(1985). 
\bibitem{r1} {\em The second DA$\Phi$NE physics handbook}, edited by
L. Maiani, G. Pancheri, and N. Paver, published 1995 by INFN-LNF-
Divisione Ricerca, ISBN 88-86409-02-8.
\bibitem{r2} F. Anulli {\em et al.}, {\em Measurement of two photon
interactions with the KLOE small angle tagging system}, p. 607 
in Vol. II of Ref.\cite{r1}. 
\bibitem{cello1} CELLO Collaboration, H.-J.Behrend {\em et al.}, 
Z. Phys. C {\bf 49}, 401(1991).
\bibitem{cello2} CELLO Collaboration, V.Savinov {\em et al.}, Report No. 
hep-ex/9507005.
\bibitem{tpc} TPC/2$\gamma$ Collaboration, H.Aihara {\em et al.}, 
Phys. Rev. Lett. {\bf 64}, 172(1990).
\bibitem{private} private communication with Prof. Nathan Isgur. 
\end{references}
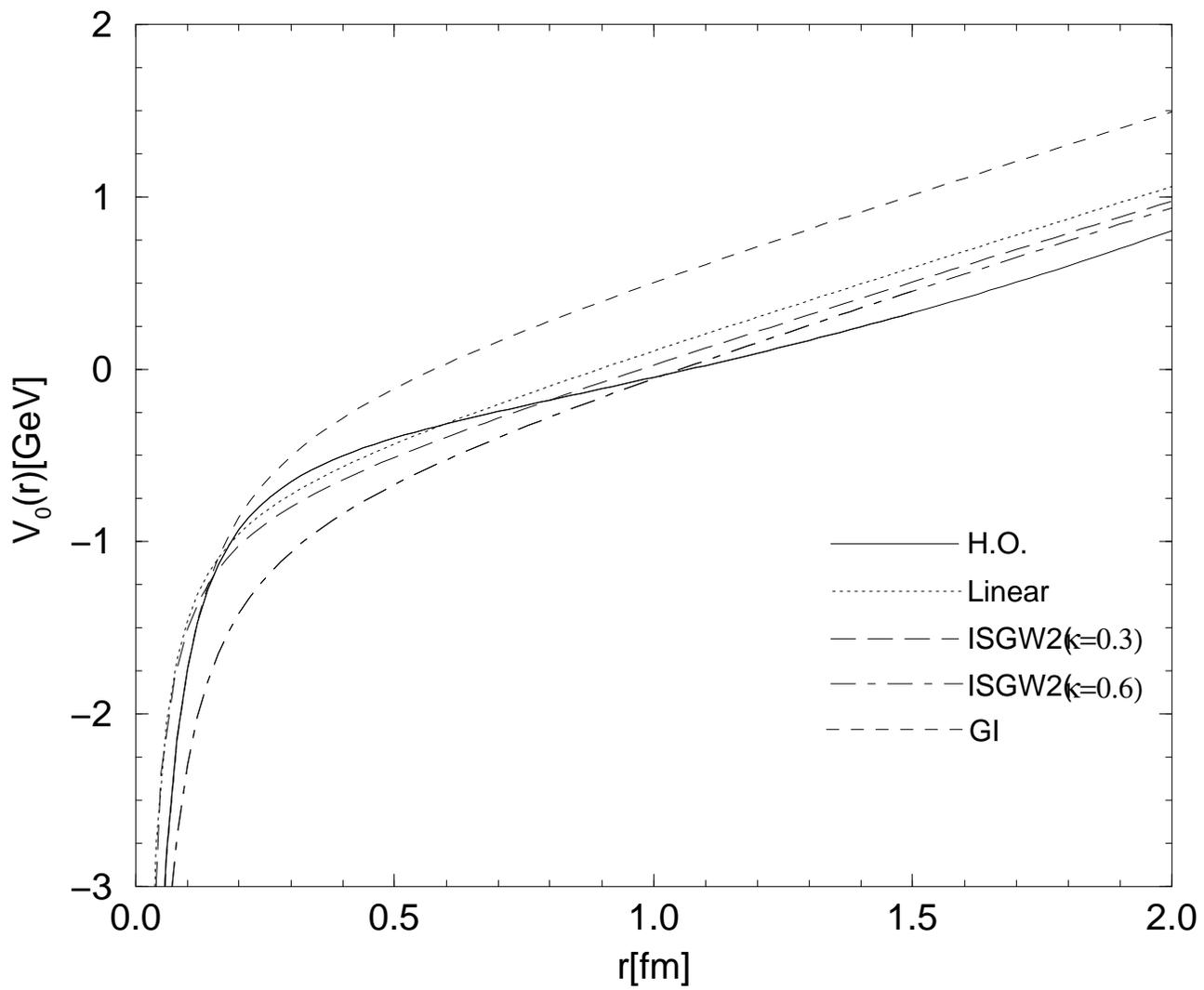
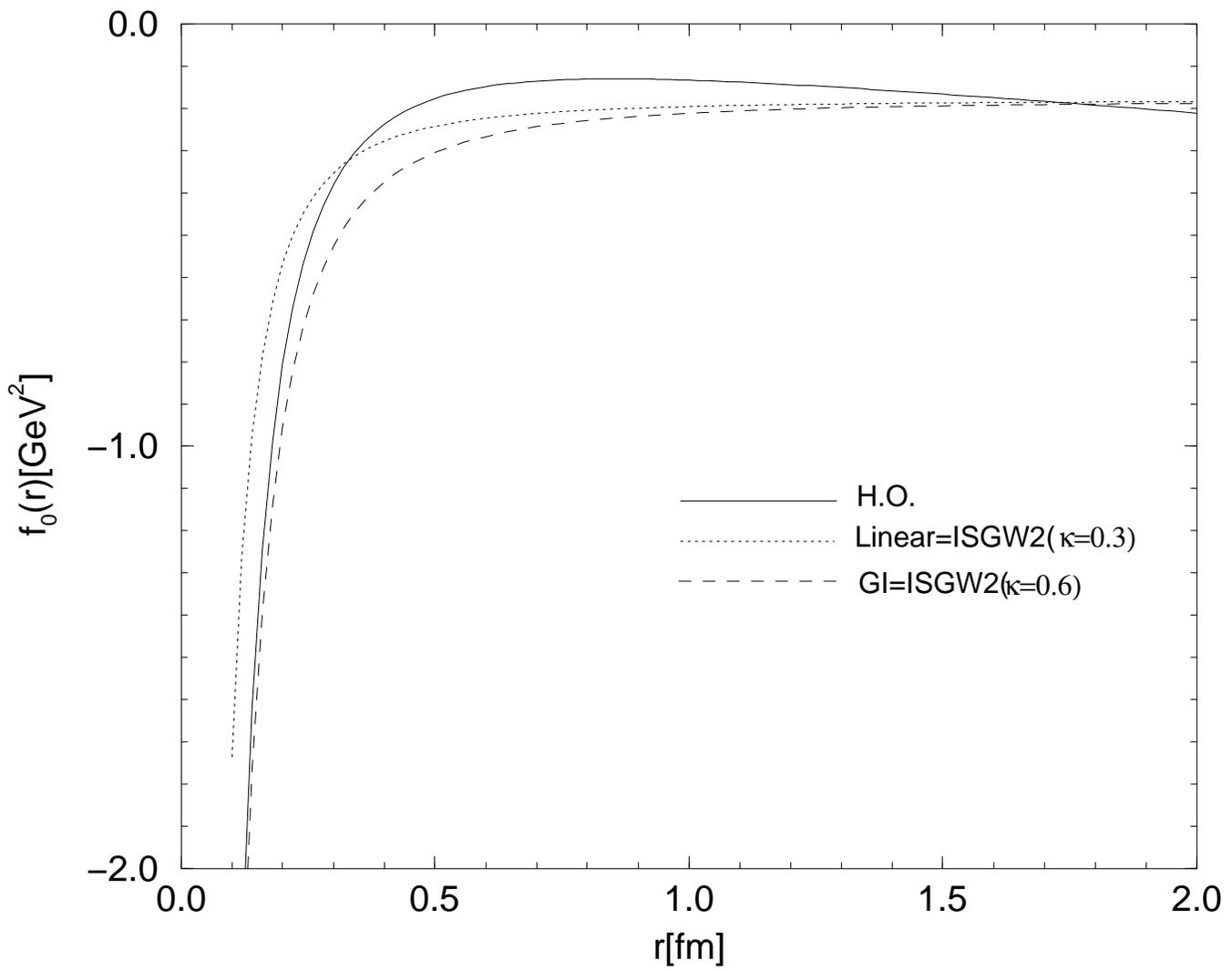
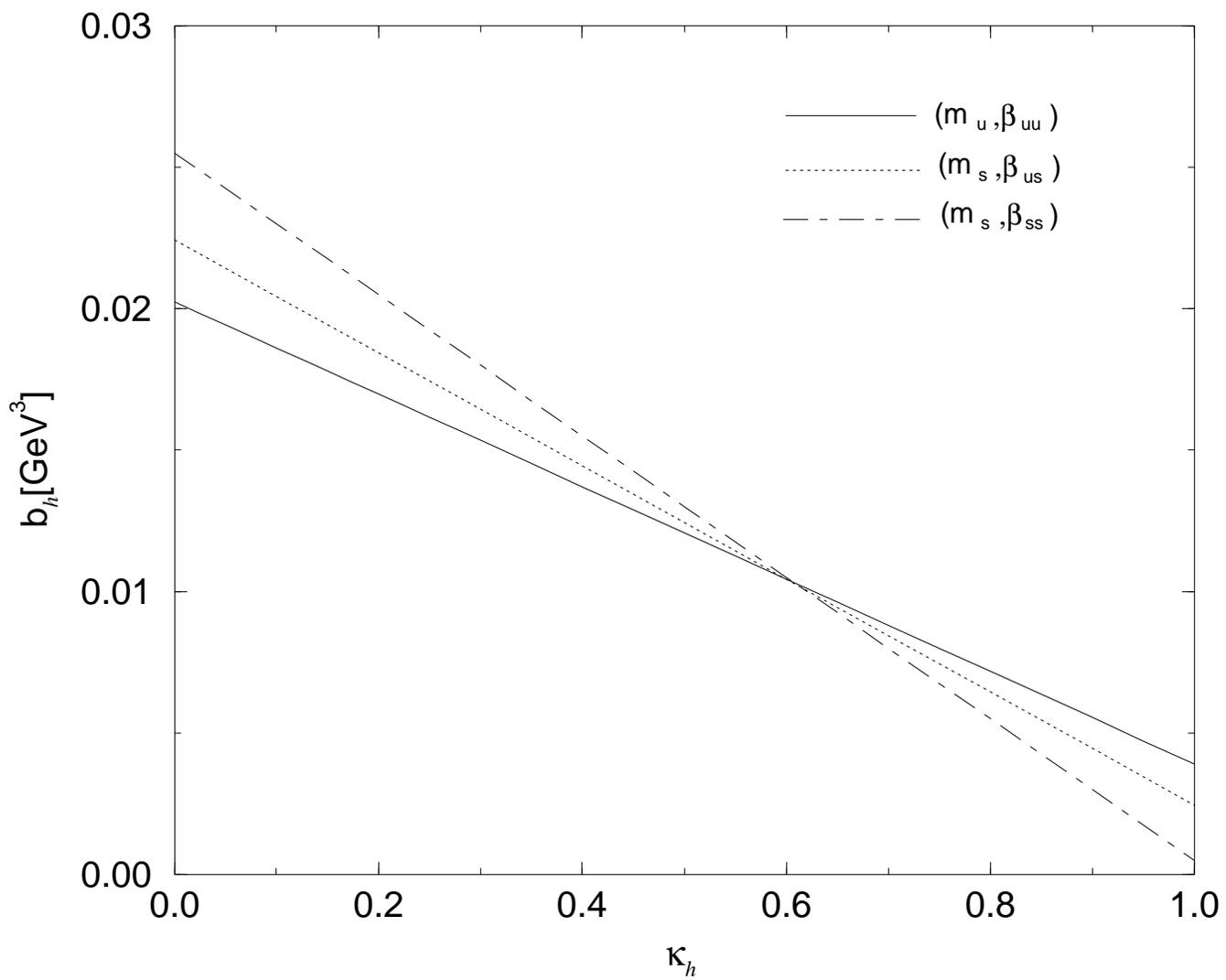
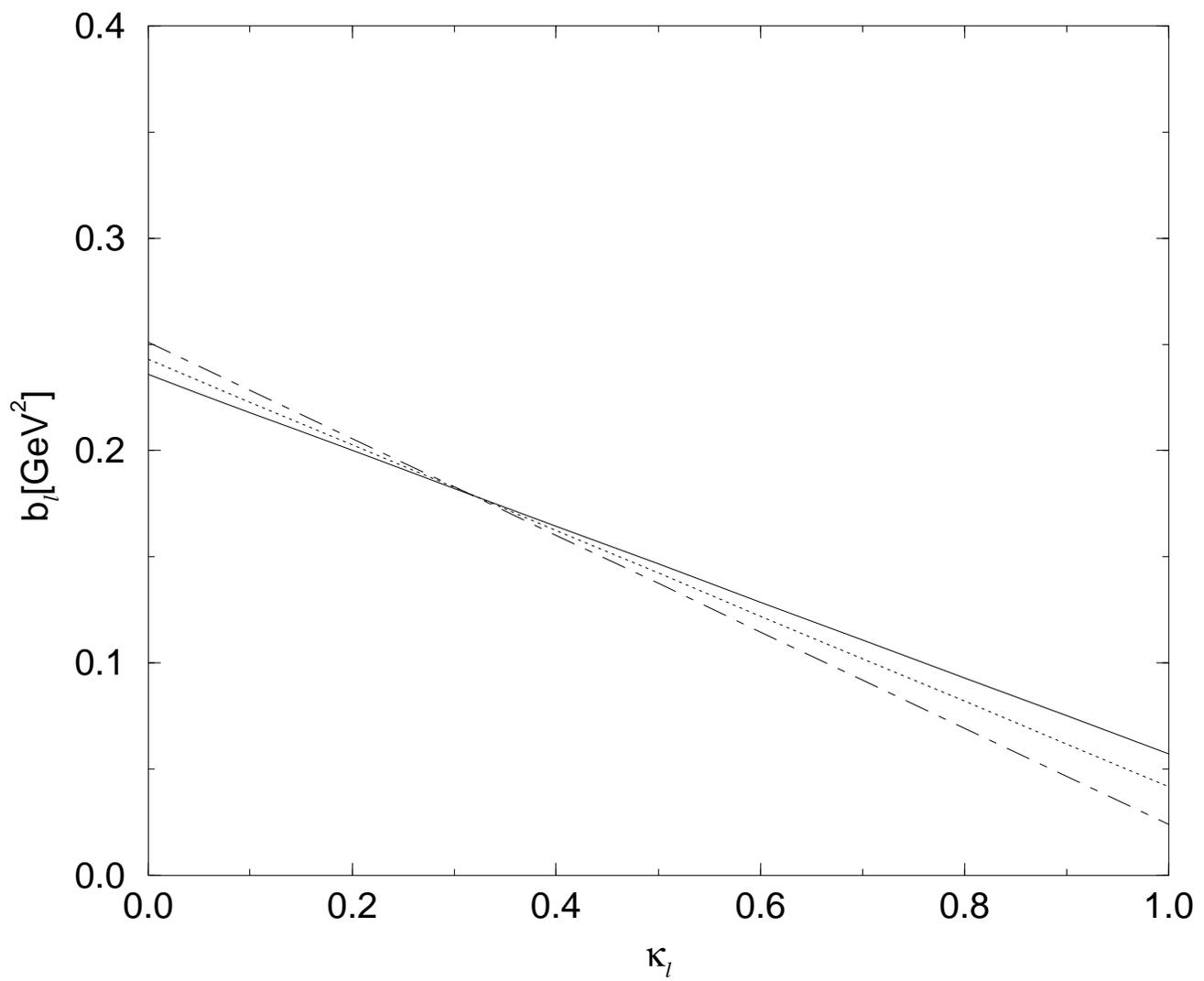
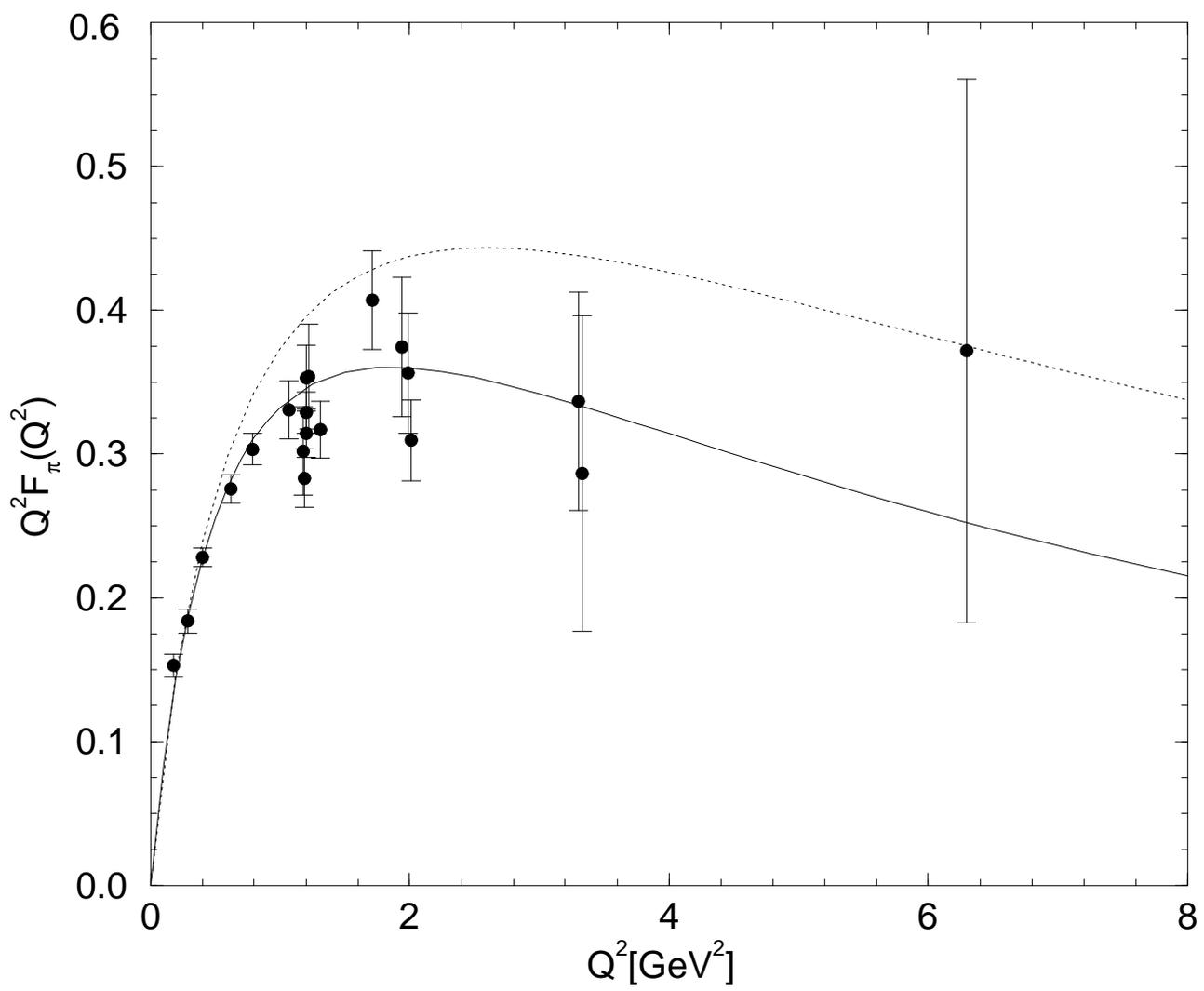
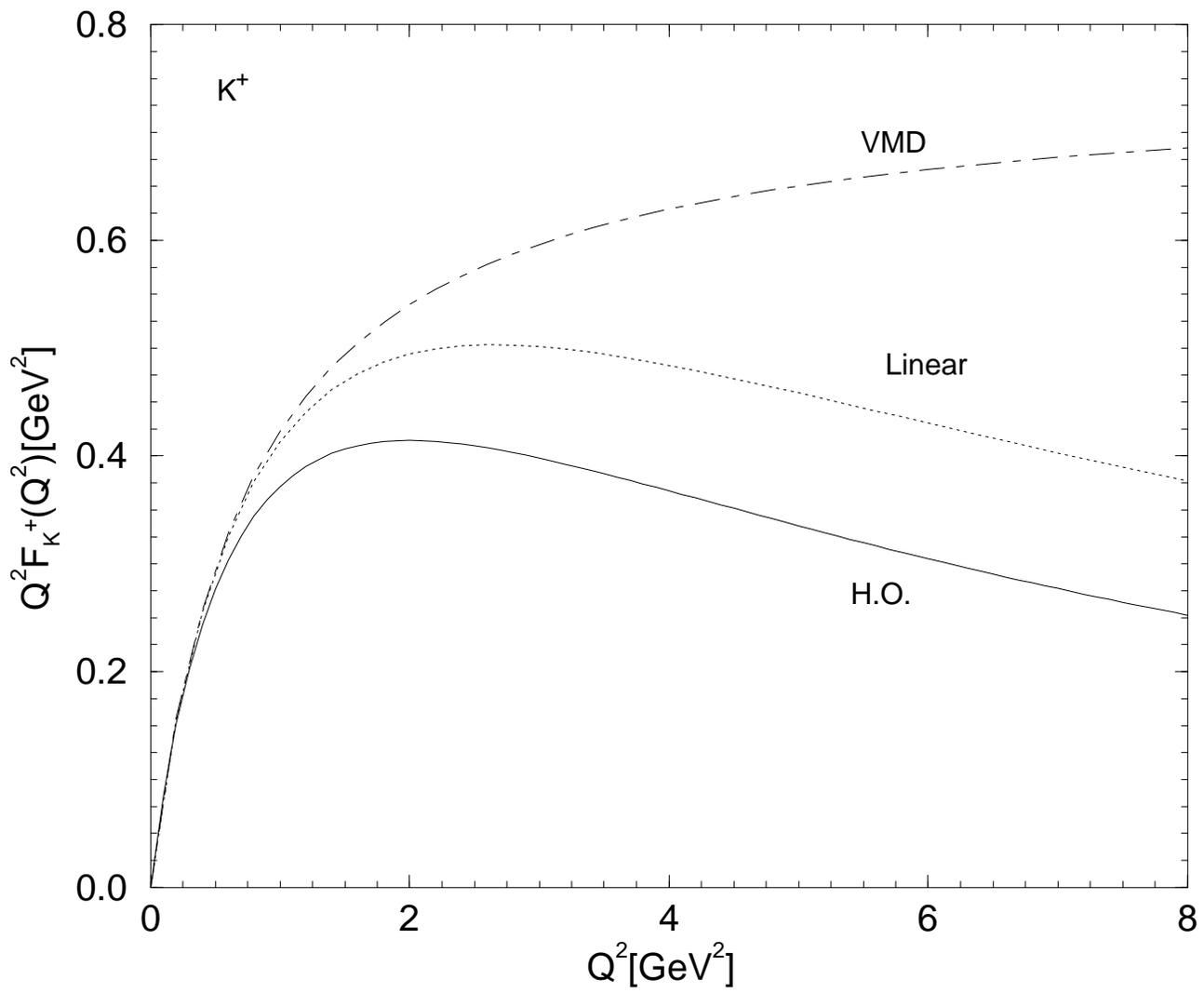
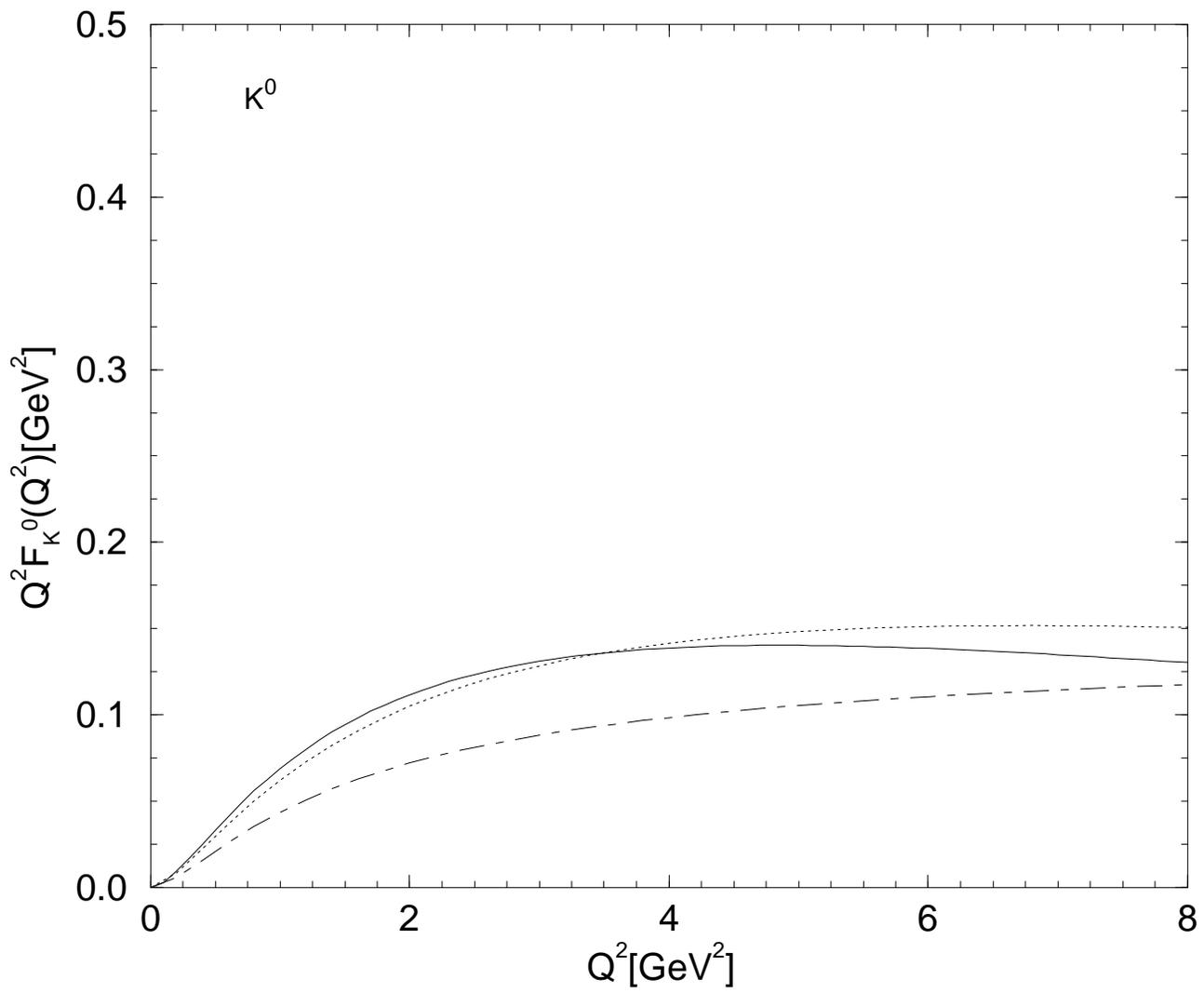
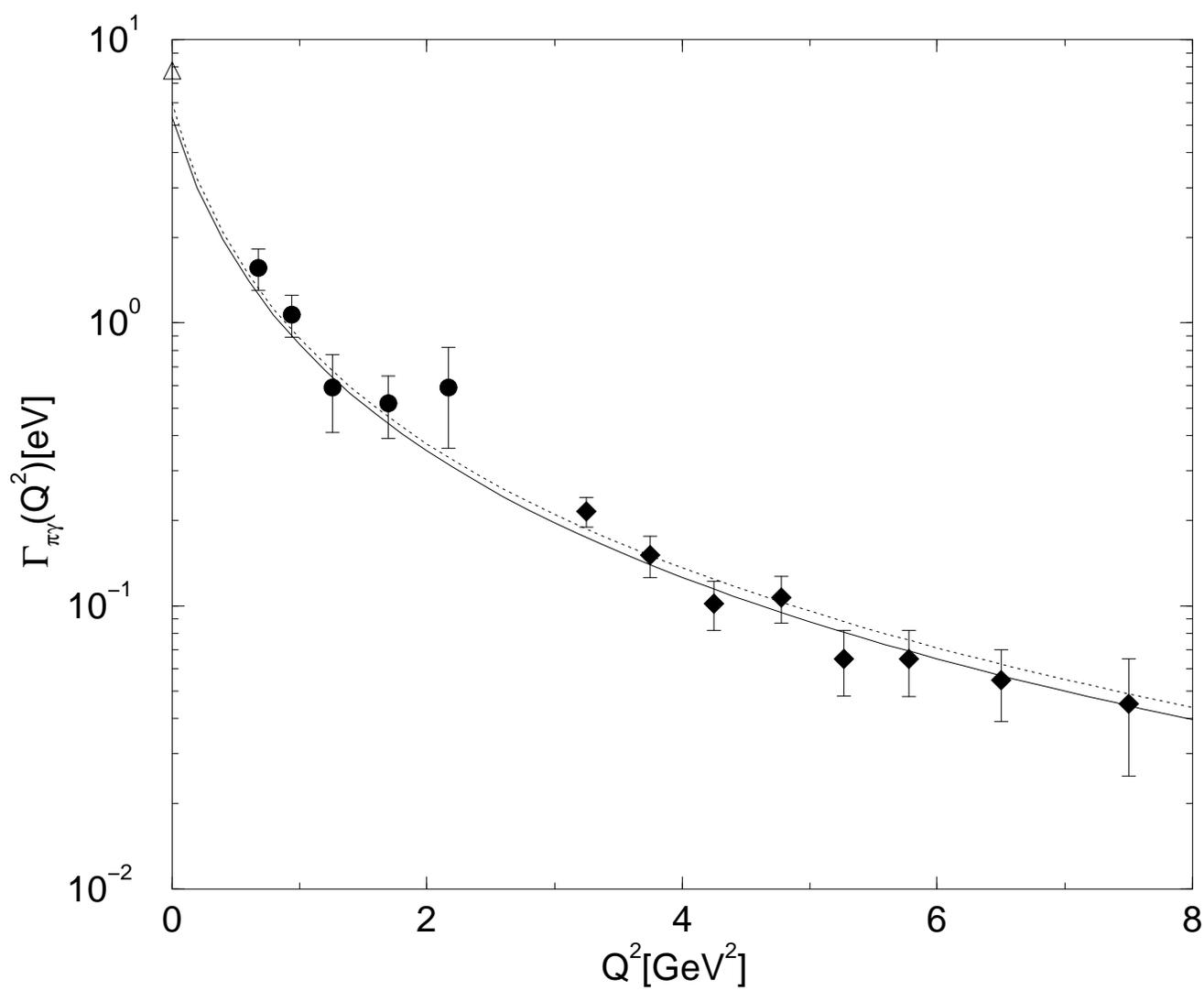
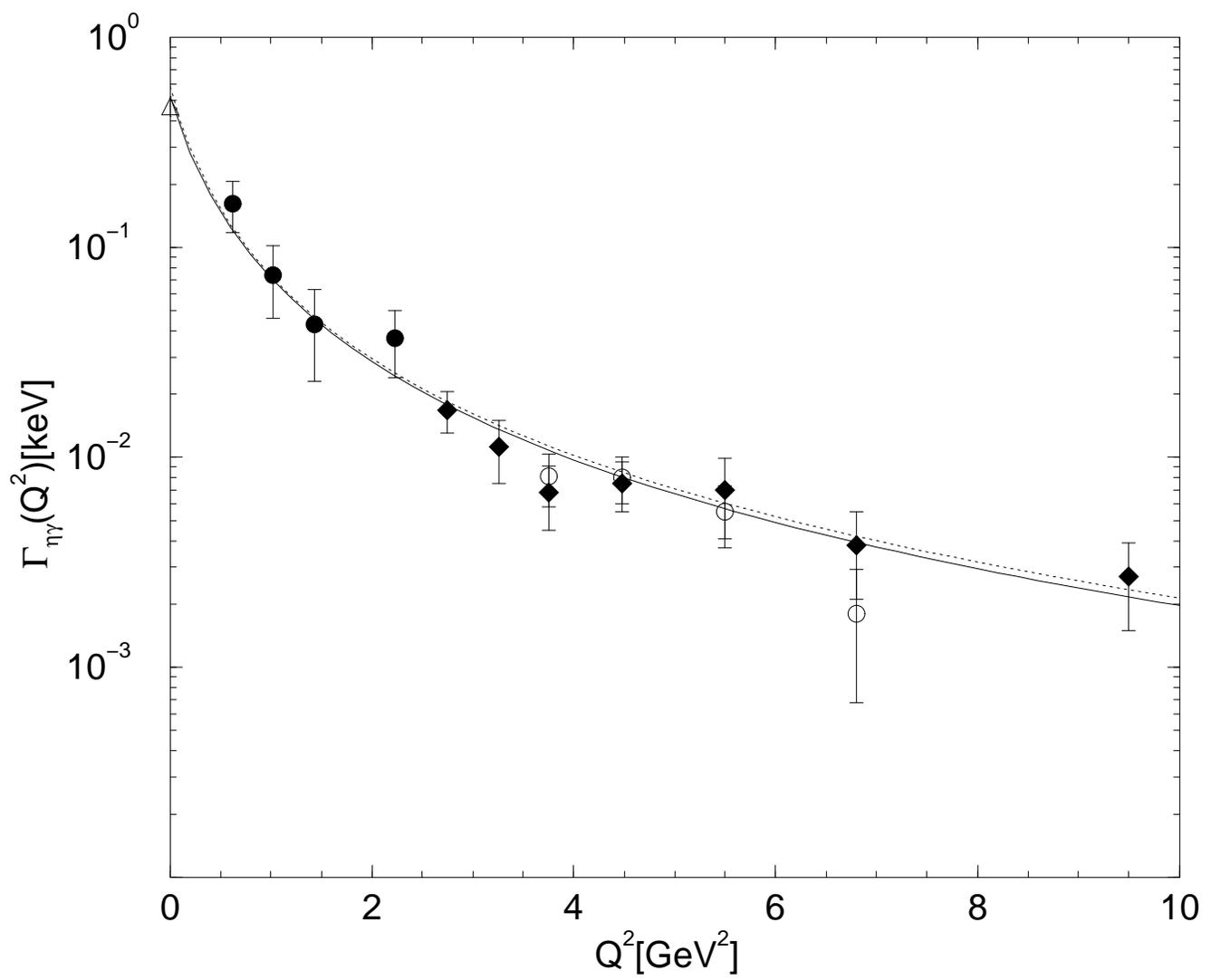
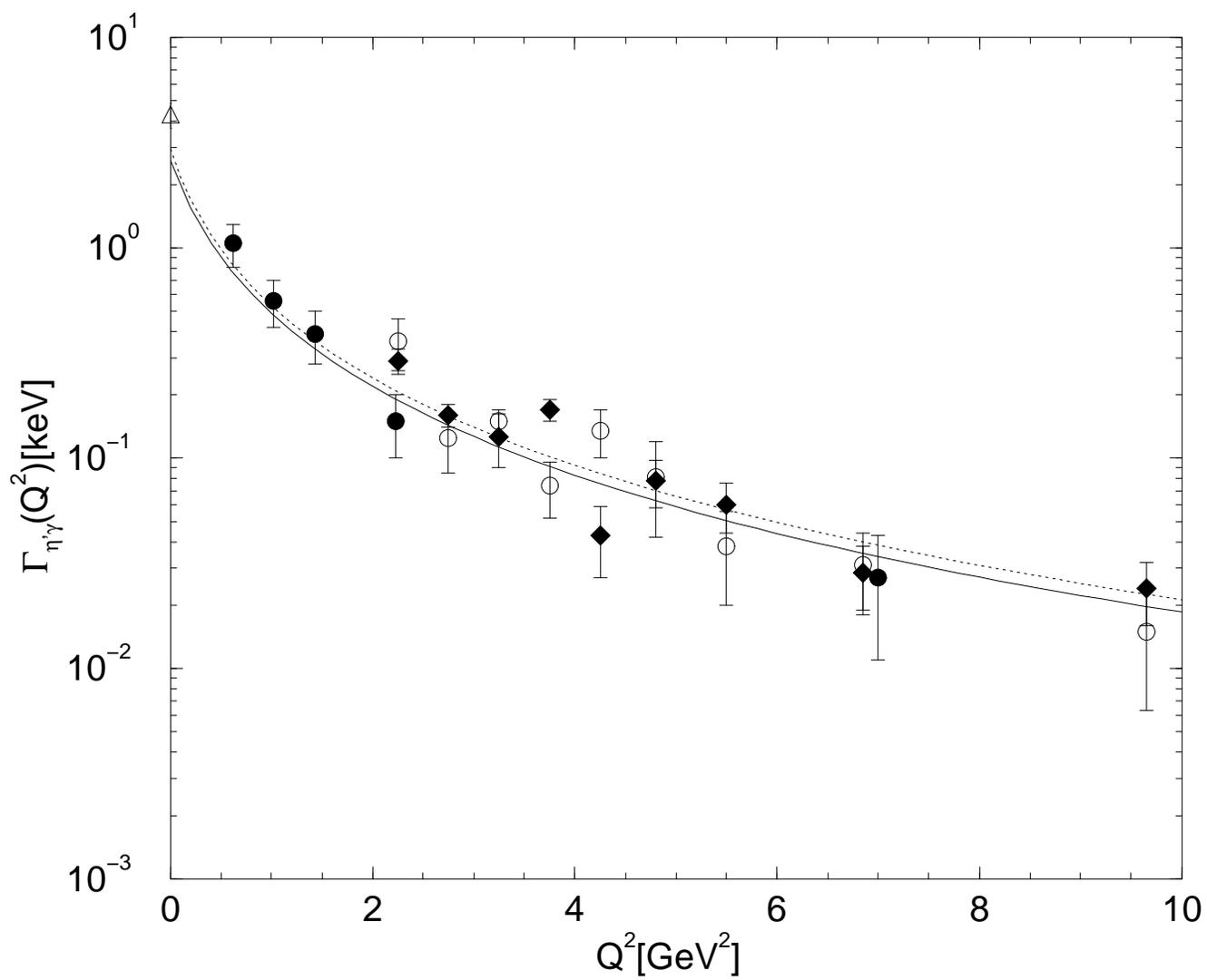
\end{document}